\documentclass[sigconf]{acmart}
\usepackage{epsfig,endnotes}
\usepackage{multirow}
\usepackage{graphicx}
\usepackage{subfig}
\usepackage{grffile}
\usepackage{thmtools, thm-restate}
\usepackage{caption}
\usepackage{color, url}
\usepackage{xspace} 
\usepackage{mathrsfs}
\usepackage{amsmath}
\usepackage{epstopdf}
\usepackage{balance}
\usepackage{bm}
\usepackage{makecell}
\usepackage{cleveref}
\usepackage{algorithm}
\usepackage{algorithmic}

\newcommand{\argmax}{\operatornamewithlimits{argmax}}
\newcommand{\myparatight}[1]{\smallskip\noindent{\bf {#1}:}~}

\graphicspath{ {figs/} }

\AtBeginDocument{%
  \providecommand\BibTeX{{%
    \normalfont B\kern-0.5em{\scshape i\kern-0.25em b}\kern-0.8em\TeX}}}

\copyrightyear{2021}
\acmYear{2021}
\acmConference[AsiaCCS '21]{2021 ACM ASIA Conference on Computer and Communications Security}{June 07--11, 2021}{Hong Kong, China}
\acmBooktitle{2021 ACM ASIA Conference on Computer and Communications Security (AsiaCCS '21), June 7--11, 2021, Hong Kong, China}
\acmPrice{15.00}
\acmDOI{10.1145/xxxxxxx.xxxxxxx}
\acmISBN{978-x-xxxx-xxxx-x/YY/MM}
\begin{document}
\date{}

\title{IPGuard: Protecting Intellectual Property of Deep Neural Networks via Fingerprinting the Classification Boundary}

\author{Xiaoyu Cao}
\affiliation{%
  \institution{Duke University}
}
\email{xiaoyu.cao@duke.edu}

\author{Jinyuan Jia}
\affiliation{%
  \institution{Duke University}
}
\email{jinyuan.jia@duke.edu}

\author{Neil Zhenqiang Gong}
\affiliation{%
  \institution{Duke University}
}
\email{neil.gong@duke.edu}

\begin{abstract}
A deep neural network (DNN) classifier  represents a model owner's intellectual property as training a DNN classifier often requires lots of resource. 
Watermarking was recently proposed to protect the intellectual property of DNN classifiers.
However, watermarking suffers from a key limitation: it sacrifices the utility/accuracy of the model owner's classifier because it tampers the classifier's training or fine-tuning process. In this work, we propose IPGuard, the first method to protect intellectual property of DNN classifiers that provably incurs no accuracy loss for the classifiers.  Our key observation is that a DNN classifier can be uniquely represented by its classification boundary. Based on this observation, IPGuard extracts some data points near the classification boundary of the model owner's classifier and uses them to fingerprint the classifier.  A DNN classifier is said to be a pirated version of the model owner's classifier if they predict the same labels for most fingerprinting data points. IPGuard is qualitatively different from watermarking. Specifically,  IPGuard \emph{extracts} fingerprinting data points near the classification boundary of a classifier that is already trained, while watermarking \emph{embeds} watermarks into a classifier during its training or fine-tuning process.  We extensively evaluate IPGuard on CIFAR-10, CIFAR-100, and ImageNet datasets. Our results show that IPGuard can robustly identify post-processed versions of the model owner's classifier as pirated versions of the classifier, and IPGuard can identify classifiers, which are not the  model owner's classifier nor its post-processed versions, as non-pirated versions of the classifier. 
\end{abstract}



\maketitle


\section{Introduction}
Suppose a model owner trains a  DNN classifier and deploys it as a cloud service or a client-side software (e.g., mobile app, Amazon Echo). 
An attacker could steal or pirate the model parameters of the classifier via malware infection, insider threats, or recent model extraction attacks~\cite{tramer2016stealing,WangHyper18,juuti2018prada,oh2017towards,hua2018reverse, yan2018cache,Hu19}. The attacker then deploys the pirated classifier as its own cloud service or client-side software. Such attacks violate the intellectual property of the model owner as training a DNN classifier often requires lots of resource, e.g., proprietary training data, confidential algorithm, and computational infrastructure.  Therefore, it is an urgent problem to protect the intellectual property of DNN classifiers. 

\begin{figure}[!t]
\centering
{\includegraphics[width=0.35 \textwidth]{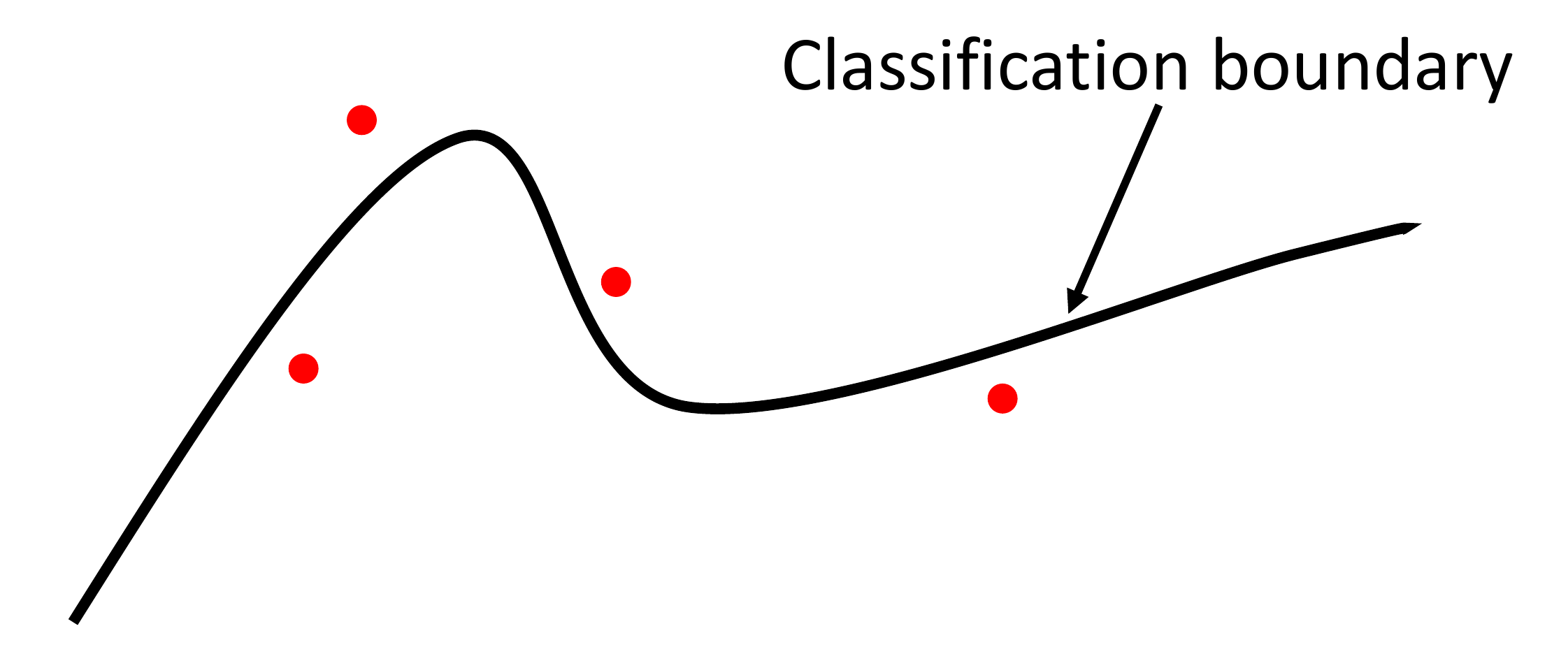}}
\caption{IPGuard leverages some data points near the target classifier's classification boundary and their labels predicted by the target classifier to fingerprint the target classifier.}
\label{illustration}
\end{figure}

Watermarking is a standard general technique to protect intellectual property and has been widely studied for multimedia data~\cite{hartung1999multimedia}. Therefore, it seems a natural solution to apply watermarking to protect intellectual property of DNN classifiers. Indeed, multiple studies~\cite{nagai2018digital, chen2018deepmarks, adi2018turning, merrer2017adversarial, darvish2019deepsigns, zhang2018protecting, guo2018watermarking} have recently extended watermarking techniques to DNN classifiers. Specifically, watermarking embeds a watermark into the model owner's classifier via modifying its training or fine-tuning process. A classifier (e.g., the classifier pirated and deployed by an attacker) is verified to be a pirated version of the model owner's classifier if  the same or similar watermark can be extracted from the classifier. For convenience, we call the model owner's classifier and the classifier to be verified \emph{target classifier} and \emph{suspect classifier}, respectively. If a suspect classifier is verified to be potentially pirated from the target classifier, then the model owner can take further follow-up actions, e.g., collecting other evidence and filing a lawsuit. 

A key limitation of watermarking is that it sacrifices the utility/accuracy of the target classifier because it tampers the training or fine-tuning process of the target classifier. In critical application domains such as healthcare and finance, even 1\% accuracy loss is intolerable.   

In this work, we propose \emph{IPGuard}, the first method to protect the intellectual property of DNN classifiers that provably incurs no accuracy loss of the target classifier. Our key observation is that a DNN classifier can be uniquely represented by its \emph{classification boundary}, which essentially divides the data domain into regions, each of which is predicted to have the same label by the classifier. Based on the observation, IPGuard finds some data points \emph{near} the target classifier's classification boundary (see Figure~\ref{illustration}). Moreover, IPGuard treats these data points (we call them \emph{fingerprinting data points}) and their labels predicted by the target classifier as a \emph{fingerprint} for the target classifier. For a suspect classifier, the model owner queries its prediction API to get the labels of the fingerprinting data points. If the suspect classifier and the target classifier predict the same labels for most fingerprinting data points, then the model owner verifies that the suspect classifier is potentially pirated from the target classifier. Unlike watermarking, IPGuard provably incurs no accuracy loss for the target classifier, because IPGuard does not tamper its training or fine-tuning process at all. 

We aim to achieve three critical goals, i.e., \emph{robustness}, \emph{uniqueness}, and \emph{efficiency}. An attacker may post-process the target classifier before deploying it, e.g., the attacker may reduce the size of the target classifier via model compression techniques before deploying it on resource-constrained devices such as mobile phone and IoT device.  Therefore, our method should be robust against post-processing, i.e., our method should still verify a suspect classifier if it is a post-processed version of the target classifier. Uniqueness means that if a suspect classifier is not the target classifier nor its post-processed version, then our method should not verify it as a pirated version of the target classifier.
Efficiency means that our method should  efficiently find the fingerprinting data points for large-scale DNN classifiers. We note that efficiency is crucial in many scenarios. For example, when the target classifier is periodically updated because the data distribution changes over time, the fingerprints need to be periodically updated. Another example is that the target classifier is trained locally on a resource-constrained IoT/mobile device, e.g., because a user does not want to share its data with a cloud server for various reasons such as privacy concerns and communication overhead. As a result, the fingerprints are also generated on resource-constrained IoT/mobile devices.

To achieve robustness and uniqueness, we aim to find fingerprinting data points \emph{near} the target classifier's classification boundary. We do not use fingerprinting data points \emph{on} the classification boundary because their predicted labels are not robust against post-processing, and we do not use fingerprinting data points that are far away from the classification boundary because their predicted labels are not unique to the target classifier and its post-processed versions. 
One way to find such fingerprinting data points  seems  to use \emph{adversarial examples} methods, i.e., treating adversarial examples as  fingerprinting data points. However, existing adversarial example methods (e.g.,~\cite{szegedy2013intriguing,goodfellow2014explaining, kurakin2016adversarial, carlini2017towards}) are insufficient because they either find adversarial examples that are far away from the classification boundary or are inefficient. For instance, FGSM~\cite{goodfellow2014explaining} can efficiently find adversarial examples but they are far away from the classification boundary.  CW~\cite{carlini2017towards} can find adversarial examples near the classification boundary but it is inefficient. The key reason is that CW aims to find human imperceptible adversarial examples. In particular, given an initial example/data point, CW aims to search for a minimum noise to turn the example into an adversarial example. Searching for such minimum noise is inefficient. 

Therefore, we propose a new method to efficiently find data points near the target classifier's classification boundary. We formulate finding such a data point as an optimization problem. Moreover, we use gradient descent to solve the problem, i.e., we start from an initial data point and iteratively move the data point along the gradient of the objective function. Our method is efficient because, unlike CW, we do not constrain the noise added to the initial data point. 

We evaluate IPGuard and compare it with multiple existing adversarial example methods on CIFAR-10, CIFAR-100, and ImageNet datasets. We consider six popular post-processing techniques including  2 variants of fine-tuning, 2 variants of retraining, and 2 variants of model compression, which an attacker may apply to a target classifier to obtain pirated classifiers. On each dataset, we also consider dozens of classifiers, which are not post-processed versions of a target classifier, as suspect classifiers. We propose a new metric to jointly measure robustness-uniqueness tradeoffs. Our results show that IPGuard can verify the post-processed versions of a target classifier as pirated versions of the classifier, and IPGuard does not falsely  verify the suspect classifiers, which are not the target classifier nor its post-processed versions, as pirated versions of the target classifier. In other words, IPGuard can achieve  robustness and uniqueness simultaneously.  Our results essentially indicate that we can find data points near a target classifier's classification boundary such that the predicted labels of these data points are robust against post-processing of the target classifier but are different for other classifiers.   
Moreover, IPGuard outperforms existing adversarial example methods. For instance, IPGuard and CW~\cite{carlini2017towards} can achieve comparable robustness-uniqueness tradeoffs, but IPGuard is orders of magnitude more efficient than CW. 

To summarize, our contributions are as follows:

\begin{itemize}

\item We propose IPGuard, the first method to protect intellectual property of DNN classifiers that provably incurs no  loss on the classifiers' accuracies. 
\item We propose a new method to efficiently find data points near a classifier's classification boundary. 
\item We propose a new metric to  measure the robustness-uniqueness tradeoffs of a fingerprinting method. Moreover, we empirically compare IPGuard with existing adversarial example methods on three datasets. 

\end{itemize}


\section{Related Work}

\subsection{Watermarking}

An attacker could steal or pirate a DNN classifier  through various methods. For instance, an insider can steal and sell the DNN classifier to a competitor. An attacker could also compromise the model owner's machine that stores the DNN classifier via social engineering attacks and malware if the machine has system and software vulnerabilities, and then steal the DNN classifier. Moreover, an attacker could also leverage recent attacks~\cite{tramer2016stealing,WangHyper18,juuti2018prada,oh2017towards,hua2018reverse, yan2018cache, Hu19} (e.g., hardware-based side-channel attacks~\cite{hua2018reverse, yan2018cache, Hu19}) that exploit the unique characteristics of DNN classifiers.

Watermarking~\cite{nagai2018digital, chen2018deepmarks, adi2018turning, merrer2017adversarial, darvish2019deepsigns, zhang2018protecting, guo2018watermarking,li2019prove} was recently proposed to protect the intellectual property of DNN classifiers. Specifically, watermarking embeds a  watermark into a target classifier during the process of training or fine-tuning the target classifier.  A suspect classifier is verified to be a pirated version of the target classifier if the same or similar watermark can be  extracted from it.   

Depending on how the watermark is embedded into a DNN classifier, we can roughly group watermarking techniques into two categories, i.e., \emph{parameter-based watermarking} and \emph{label-based watermarking}. In parameter-based watermarking~\cite{nagai2018digital, chen2018deepmarks},  the watermark is embedded into the model parameters of the target DNN classifier via adding new regularization terms to the loss function during training. Parameter-based watermarking requires white-box access to the model parameters of the suspect classifier in order to verify the watermark. In label-based watermarking~\cite{adi2018turning, merrer2017adversarial, darvish2019deepsigns, zhang2018protecting, guo2018watermarking,li2019prove}, the watermark is embedded into the predicted labels or neuron activations of certain data points. Specifically, they first pick some data points (e.g., abstract images~\cite{adi2018turning}, training data points with extra meaningful content~\cite{zhang2018protecting},or adversarial examples \cite{merrer2017adversarial,li2019prove}) and assign certain labels to them. Then, they use these data points to augment the training data to learn the target classifier. Given a suspect classifier, they query the suspect classifier to obtain labels of the data points. If labels predicted by the suspect classifier match those predicted by the target classifier, then the watermark is verified. 

\myparatight{Limitations}  Watermarking techniques suffer from two key limitations. First, they inevitably sacrifice utility/accuracy of the target classifier because they require tampering the target classifier's training or fine-tuning process. It is a well known challenge to search for a good DNN architecture  that has a high accuracy for a particular task. For instance, on the ImageNet dataset, it took lots of trial-and-error efforts for the computer vision community to search for the DNN architecture called ResNet152V2~\cite{he2016identity}, which increases the testing accuracy upon the DNN architecture called ResNet152~\cite{he2016deep} by 1\%~\cite{chollet2015keras}. However, watermarking~\cite{darvish2019deepsigns} easily decreases the testing accuracy by 0.5\% when embedding only 20 watermark data points into an ImageNet model. 
Second, watermarking techniques cannot be applied to legacy target classifiers that cannot be retrained because they need to \emph{embed} watermarks by tampering the training process. Instead, in this work, we propose a method that \emph{extracts} adversarial examples from the target classifiers as fingerprinting data points without changing the training process, which is applicable to legacy target classifiers.

\subsection{Adversarial Examples}
\label{sec:adv_example}

Suppose we are given an example/data point and a classifier, where the classifier can correctly predict the label of the example. We can add some carefully crafted noise to the  example such that the classifier predicts a label as we desire. The  example with carefully crafted noise is called an \emph{adversarial example}~\cite{szegedy2013intriguing}. 
In particular, we may desire the classifier to predict any incorrect label for the adversarial example (called \emph{untargeted adversarial example}) or predict a particular incorrect label for the adversarial example (called \emph{targeted adversarial example}). 
Many methods (e.g.,~\cite{goodfellow2014explaining, kurakin2016adversarial, carlini2017towards}) have been developed to construct adversarial examples.
 
An adversarial example essentially moves an example across the classifier's classification boundary via adding carefully crafted noise to the example. Therefore, it seems a natural choice to use adversarial examples to fingerprint a classifier's classification boundary. 
Therefore, we review several popular methods~\cite{goodfellow2014explaining, kurakin2016adversarial, carlini2017towards} to construct adversarial examples. We focus on targeted adversarial examples as we aim to construct adversarial examples for each class to better characterize the classifier's classification boundary. 

\myparatight{Fast Gradient Sign Method (FGSM)} FGSM~\cite{goodfellow2014explaining} is an efficient method to construct adversarial examples. Given an initial example/data point $x$ and a noise bound $\epsilon$, FGSM returns the following $x'$ as an adversarial example:
\begin{align}
x' = \text{clip}(x - \epsilon\cdot \text{sign}(\nabla_x J(C, y^*; x))),
\end{align}
where $C$ is the classifier, $J$ is the loss function used to train $C$, $y^*$ is the target label we desire for the adversarial example, $\nabla_x$ represents gradient, and $\text{clip}$ is the function to project the adversarial example back to the feasible data domain, e.g., each pixel is projected back to be in the range [0, 1] if the example is an image. 

\myparatight{Iterative Gradient Sign Method (IGSM)} IGSM ~\cite{kurakin2016adversarial} is an improved version of FGSM. Instead of taking just one single step, IGSM moves the example iteratively, one small step at a time. In the $t$th iteration, IGSM updates the example with the following rule:
\begin{align}
x'_t = \text{clip}(x'_{t-1} - \text{clip}_\epsilon (\alpha\cdot \text{sign}(\nabla_x J(C, y^*; x)))), 
\end{align}
where $\alpha$ is a small step size and $\text{clip}_\epsilon(z)$ is the function to project each dimension of $z$ to be at most $\epsilon$.  
The process is repeated until a successful adversarial example is generated or the maximum number of iterations is reached. 

\myparatight{Carlini and Wagner's methods (CW)} CW methods~\cite{carlini2017towards} have three variants depending on which metrics are used to measure the noise added to adversarial examples. Specifically,  CW-$L_0$, CW-$L_2$, and CW-$L_\infty$ use $L_0$-norm, $L_2$-norm, and $L_\infty$-norm to measure the magnitude of the noise, respectively. We consider the CW-$L_2$ method as it is much more efficient than CW-$L_0$ and CW-$L_\infty$. In particular,  given an initial example, CW-$L_2$ iteratively searches for a small noise that turns the example into an adversarial example.  

\myparatight{Limitations} Existing adversarial example methods either achieve suboptimal robustness-uniqueness tradeoffs (we will show experimental results) or are inefficient. Specifically, FGSM and IGSM achieve suboptimal robustness-uniqueness tradeoffs. A key reason is that the predicted labels of their constructed adversarial examples are either not robust against post-processing of the target classifier or not unique to the target classifier.  CW-$L_2$ achieves comparable robustness-uniqueness tradeoffs with our IPGuard. However,  CW-$L_2$ is inefficient because it aims to find adversarial examples with small noise. For instance, CW-$L_2$ takes around 20 mins to find one adversarial example on an ImageNet classifier, while our method only takes around 76 seconds to find one data point near the classification boundary for the same classifier. 


\section{Problem Definition}

\subsection{Threat Model}\label{sec:threat_model}
We consider two parties, i.e., \emph{model owner} and \emph{attacker}. A model owner trains a DNN classifier (called \emph{target classifier}) using a (proprietary) training dataset and algorithm.  The model owner could deploy the target classifier as a cloud service (also known as \emph{machine learning as a service}) or as a client-side software (e.g., Amazon Echo).  
 An attacker pirates the target classifier and deploys it as its own software or service. Moreover, the attacker may post-process the target classifier before deploying it. For example, to deploy the classifier on resource-constrained devices such as smartphone and IoT device, the attacker may use model compression methods (e.g., pruning)~\cite{han2015learning, li2016pruning} to reduce the size of the classifier.    

The model owner derives a \emph{fingerprint} for its target classifier. For a {suspect classifier}, the model owner verifies whether it has the same fingerprint via leveraging its prediction API. If the suspect classifier has the same fingerprint, then the suspect classifier is likely pirated from the model owner's target classifier and the model owner can take further follow-up actions, e.g., collecting other evidence and filing a lawsuit.      

\subsection{Fingerprinting a DNN Classifier}
We define fingerprinting a DNN classifier as designing two functions, i.e., \emph{Extract} and \emph{Verify}. We describe the two functions as follows:

\myparatight{Extract} Given a target classifier $C_t$, the model owner executes the Extract function to derive a fingerprint as $f_{C_t}=Extract(C_t)$, where $f_{C_t}$ is the fingerprint.

\myparatight{Verify} Given a suspect classifier $C_s$, $Verify(f_{C_t}, C_s)$ produces either 1 or 0, where 1 means that the suspect classifier is predicted to be a pirated version of the target classifier. We consider that the Verify function only leverages the suspect classifier's prediction API, which is generally applicable no matter whether the suspect classifier is deployed as a client-side software or a cloud service. Moreover, we assume the prediction API returns a predicted label for a query. We note that some prediction APIs return a confidence score vector for a query, where the confidence score vector is a probability distribution of the query's label.  However, we only use the predicted label in this work and leave leveraging the confidence score vector to fingerprint a classifier as future work.  

\myparatight{Difference with watermarking}  Fingerprinting a DNN classifier  is \emph{qualitatively} different from watermarking a DNN classifier. Specifically, watermarking  is to \emph{embed} a  watermark into a classifier during the process of training or fine-tuning the classifier, while fingerprinting is to \emph{extract} a fingerprint from a classifier that is already trained. Therefore, watermarking inevitably sacrifices the classifier's accuracy as it tampers the training or fine-tuning process. For the same reason, watermarking is not applicable to a legacy classifier that is challenging to re-train.

\subsection{Design Goals}
\label{property}
We aim to design a fingerprinting method that has the following properties. 
\begin{itemize}
\item \textbf{Fidelity}. The fingerprinting method should  not sacrifice  the target classifier's accuracy at all. Watermarking methods do not have such property. 
\item \textbf{Effectiveness}. If the suspect classifier is the same as the target classifier, then the Verify function should produce 1. Formally, we should have $Verify(Extract(C_t), C_s)=1$ when $C_s=C_t$. 
\item \textbf{Robustness}. If the suspect classifier is some post-processed version of the target classifier (we consider popular post-processing methods that were also considered in previous work~\cite{adi2018turning,darvish2019deepsigns} and details are shown in Section~\ref{sec:susp_cls}), then the Verify function should produce 1. Formally, we should have $Verify(Extract(C_t), C_s)=1$ if $C_s$ is a post-processed version of $C_t$.  
\item \textbf{Uniqueness}. The fingerprint should be unique to the target classifier. In other words, if a suspect classifier is not the target classifier nor its post-processed version, then the Verify function should produce 0. Formally, we should have $Verify(Extract(C_t), C_s)=0$ if $C_s$ is neither $C_t$ nor a post-processed version of $C_t$.  
\item \textbf{Efficiency}. The fingerprinting method should be efficient to extract a fingerprint for a target classifier and verify the fingerprint for a suspect classifier.  
\end{itemize}

 \section{Our IPGuard}

\subsection{Overview}

A DNN classifier is essentially characterized by a \emph{classification boundary}. Different classifiers have different classification boundaries. Therefore, we aim to fingerprint the classification boundary of the target classifier. In particular, we treat a set of data points (called \emph{fingerprinting data points}) near the target classifier's classification boundary and their labels predicted by the target classifier as the fingerprint. Given a suspect classifier, the model owner queries its prediction API for the labels of the fingerprinting data points. If the predicted labels match those in the fingerprint, then  the suspect classifier is predicted to be the target classifier or a post-processed version of the target classifier. 
Specifically, we overview our Extract and Verify functions as follows:

\myparatight{Extract} Given a target classifier, our Extract function searches $n$ random data points near the classification boundary of the target classifier. We treat the $n$ data points and their labels predicted by the target classifier as the fingerprint. Formally, we have the fingerprint as $f_{C_t}=\{(x_1, y_1), (x_2, y_2), \cdots, (x_n, y_n)\}$, where   $x_i$ and $y_i$ are the $i$th fingerprinting data point and its label predicted by the target classifier. We formulate finding a data point near a classifier's classification boundary as an optimization problem and leverage gradient descent to solve the optimization problem. 

\myparatight{Verify} Given a suspect classifier, the model owner queries the classifier's prediction API for the labels of the $n$ fingerprinting data points. In particular, we denote by $y_i'$ the label predicted by the suspect classifier for the $i$th fingerprinting data point $x_i$. Suppose $m$ data points have the same labels predicted by the suspect classifier and the target classifier, i.e., $y_i'=y_i$ for $m$ data points. We define $\frac{m}{n}$ as the \emph{matching rate}. If the matching rate is larger than or equal to  a  threshold, then the Verify function produces 1, otherwise the Verify function produces 0. 

Next, we describe our Extract function to search for data points near the target classifier's classification boundary. 

\subsection{Extract}

\myparatight{Classification boundary}
Suppose the target classifier is a $c$-class DNN classifier, where the output layer is a softmax layer. Moreover, we denote by $\{g_1,g_2,\cdots,g_c\}$ the decision functions of the target classifier, i.e., $g_i(x)$ is the probability that the example $x$ has a label $i$, where $i=1, 2,\cdots,c$. For convenience, we denote by $\{Z_1,Z_2,\cdots,Z_c\}$ the logits of the target classifier, i.e., $\{Z_1,Z_2,\cdots,Z_c\}$ are the outputs of the neurons in the second-to-last layer. Formally, we have:
\begin{align}
 g_i(x)=\frac{exp(Z_i(x))}{\sum_{j=1}^cexp(Z_j(x))},
 \end{align}
where $i=1, 2,\cdots,c$. The label $y$ of the example $x$ is predicted as the one that has the largest logit or probability, i.e., $y=\argmax_i g_i(x)=\argmax_i Z_i(x)$. 

A data point is on the target classifier's classification boundary if the target classifier cannot decide the label of the data point. In other words, if at least two labels have the largest probability (or logit) for a data point, then the data point is on the target classifier's classification boundary. Formally, we can define the target classifier's classification boundary as the following set of data points:  
\begin{align}
&\text{\bf Classification boundary: } \nonumber\\
 CB&=\{x|\exists i, j, i\neq j \text{ and } g_i(x)=g_j(x)\geq \max_{t\neq i,j}g_t(x) \} \nonumber \\
 &=\{x|\exists i, j,  i\neq j \text{ and } Z_i(x)=Z_j(x)\geq \max_{t\neq i,j}Z_t(x) \},
 \end{align}
where $CB$ is the set of data points that constitute the target classifier's classification boundary. 

\myparatight{Finding data points near the classification boundary} 
 We describe how to find one data point near the target classifier's classification boundary and our method repeatedly finds   $n$ data points. 
One naive method to find a data point on the classification boundary is to repeatedly randomly sample data points and check whether they are in the set $CB$. Specifically, if a data point $x$ satisfies the condition $\exists i, j, i\neq j \text{ and } g_i(x)=g_j(x)\geq \max_{t\neq i,j}g_t(x)$ or $\exists i, j, i\neq j \text{ and } Z_i(x)=Z_j(x)\geq \max_{t\neq i,j}Z_t(x)$, then the data point is on the classification boundary. However, such naive sampling method would take a very long time (if possible) to find one data point on the classification boundary. This is because the size of the set $CB$ is a negligible portion of the data domain.  

Therefore, to address the computational challenge, we  formulate finding  data points near the classification boundary as an optimization problem. Specifically, we aim to solve the following optimization problem:  
\begin{align}
\min_{x} \; &ReLU(Z_i(x)-Z_j(x)+k) \nonumber \\
+ &ReLU(\max_{\substack{t\neq i,j}}{Z_t(x)} - Z_i(x)), \label{eq:bound_k}
\end{align}
where $i$ and $j$ are randomly sampled labels, $ReLU$ is defined as $ReLU(s)=\max\{0, s\}$, and $k$ is related to the distance between the data point $x$ and the classification boundary. Intuitively, the objective function is small if $Z_j(x) \geq Z_i(x) + k$ ($Z_j(x) = Z_i(x)$ means that $x$ is on the classification boundary) and $\max_{\substack{t\neq i,j}}{Z_t(x)} \leq Z_i(x)$. The parameter $k$ balances between the robustness and uniqueness (please refer to Section~\ref{property} for details of robustness and uniqueness) of our fingerprinting method. Specifically, on one hand, when $k$ is larger,  a post-processed  version of the target classifier is more likely to predict the same label for the data point, which means that our method is more robust against post-processing. On the other hand, when $k$ is larger, a suspect classifier that is not a post-processed version of the target classifier is also more likely to predict the same label for the data point, which means that our fingerprint is less unique. We will study the impact of $k$ on the robustness-uniqueness tradeoff in experiments. 

To find a data point $x$ near the classification boundary, we solve the optimization problem in Equation~\ref{eq:bound_k} with randomly sampled $i$ and $j$. Specifically, we use the Adam optimizer~\cite{kingma2014adam} to solve the optimization problem and $x$ is initialized to be a data point whose label is predicted as $i$. We will discuss more details on how to initialize $x$ and sample $i$ and $j$ in Section \ref{sec:imp_details}.   We stop the iterative process  when the objective function equals 0 or we have  reached the maximum number of iterations (e.g., 1,000 in our experiments). Note that, when finding $n$ data points, we randomly sample different initialization, $i$, and $j$, in order to have a good coverage of the classification boundary. 


\section{Experiments}
\subsection{Experimental Setup}

\begin{table}[!t]\renewcommand{\arraystretch}{1.00}
\centering
\caption{Datasets, target classifiers, and suspect classifiers. }
\centering
\scalebox{1.0}{
\begin{tabular} {|c|c|c|c|}\hline 
{\small Dataset} & {\small Target classifier} & {\makecell{\small \# Positive\\ \small suspect \\ \small classifiers}} & {\makecell{\small \# Negative\\ \small suspect \\ \small classifiers}}\\ \hline
{\small CIFAR-10} & {\small ResNet20} & {\small 15} & {\small 80}\\ \hline
{\small CIFAR-100} &  {\small WRN-22-4} & {\small 14} & {\small 80}\\ \hline
{\small ImageNet} & {\small ResNet50} & {\small 10} & {\small 10}\\ \hline
\end{tabular}
}
\label{tab:datasets}
\end{table}

\subsubsection{Datasets and Target Classifiers}
We consider three image classification datasets that are widely used in the literature, i.e., CIFAR-10, CIFAR-100, and ImageNet. Table~\ref{tab:datasets} shows the three datasets and their target classifiers. 

\myparatight{CIFAR-10} CIFAR-10~\cite{krizhevsky2009learning}  consists of 60000 images with size $32\times32\times3$. There are 50000 images in the training set and 10000 images in the testing set. CIFAR-10 is a 10-class classification problem. 
We train a widely used ResNet20~\cite{he2016deep} model as our target classifier. 

\myparatight{CIFAR-100} Like CIFAR-10, CIFAR-100~\cite{krizhevsky2009learning}  has 60000 images and each image has a size of $32\times32\times3$. However, CIFAR-100 is a 100-class classification problem. 
We train a wide residual network WRN-22-4~\cite{zagoruyko2016wide} as our target classifier, which is an improved version of ResNet20 for CIFAR-100. 

\myparatight{ImageNet} ImageNet ~\cite{ILSVRC15}  contains about 1.2 million training examples, 50,000 validation examples, and 100,000 testing examples from 1000 classes. We consider each image is resized to $224\times 224\times 3$ and we use the pre-trained ResNet50~\cite{he2016deep} model from Keras~\cite{chollet2015keras} as our target classifier. 

\subsubsection{Suspect Classifiers}\label{sec:susp_cls} Given a dataset and a target classifier, we consider multiple categories of suspect classifiers.  

\myparatight{Post-processed versions of the target classifier} Like previous studies on watermarking~\cite{adi2018turning,darvish2019deepsigns}, we consider six post-processing including 2 variants of fine-tuning, 2 variants of retraining, and 2 variants of model compression. We expect the suspect classifiers from these post-processing to be verified by our fingerprinting method. 
\begin{itemize}
	\item{\textbf{Fine-tune last layer (FTLL)}}. Only the last layer (fully connected layer) of the target classifier is fine-tuned using the training data.  All other layers are fixed. 
	\item{\textbf{Fine-tune all layer (FTAL)}}. All layers in the target classifier are fine-tuned with the training data.
	\item{\textbf{Retrain last layer (RTLL)}}. The weights of the last layer are re-initialized and then trained with the training data. All other layers are fixed.
	\item{\textbf{Retrain all layers (RTAL)}}.  The weights of the last layer are re-initialized but the weights of all layers are retrained with the training data.
	\item{\textbf{Weight pruning (WP)}}. WP is a popular method to compress a neural network model~\cite{han2015learning}. In particular, we prune $p$ fraction of weights that have the smallest absolute values and then retrain the pruned classifier to enhance the classification accuracy. We increase $p$ from 0.1 with a step size 0.1 until the accuracy loss is larger than 3\% (model compression aims to compress a model without substantially sacrificing the model's accuracy); and each $p$ corresponds to a pruned version of the target classifier. Therefore, we may have multiple pruned target classifiers as suspect classifiers.  For instance, if we can prune at most $p=0.5$ fraction of weights from a target classifier such that the accuracy loss is  no larger than 3\%, then we have 5 suspect classifiers, which correspond to $p=0.1, 0.2, 0.3, 0.4$, and 0.5, respectively. Note that FTLL, FTAL, RTLL, or RTAL has just one post-processed version of the target classifier. 
	\item{\textbf{Filter pruning (FP)}}. FP is another popular method to compress a neural network model~\cite{li2016pruning}. In particular, we prune $c$ fraction of filters in each layer that have the smallest absolute values and then retrain the pruned classifier to enhance the classification accuracy. As shown in \cite{li2016pruning}, it can be too complicated to prune the last layer of each residual block for ResNet, so we only consider other layers in each residual block. We increase $c$ from $\frac{1}{16}$ with a step size $\frac{1}{16}$ until the accuracy loss is larger than 3\%; and each $c$ corresponds to a suspect classifier. We use the step size $\frac{1}{16}$ because the number of filters in our target classifiers can be divided by 16. 	Like WP, FP may lead to multiple suspect classifiers for a target classifier, where each suspect classifier corresponds to a value of $c$.
\end{itemize}

\begin{table}[!t]\renewcommand{\arraystretch}{1.0}
\centering
\caption{Testing accuracies of the  target and suspect classifiers. ``--'' means not applicable. If there are multiple classifiers from the same type, we show the range of accuracies.  For instance, for the same-architecture neural network classifier, we trained 50 ResNet20 models for the CIFAR-10 dataset, and their accuracies are in the range [0.91,0.92]. For positive suspect classifiers based on WP (or FP), we show the accuracies of the suspect classifiers with $p$ fraction of weights (or $c$ fraction of filters) pruned for all $p$ (or $c$) that  sacrifice the target classifier's accuracy by at most 3\%.}
\centering
\addtolength{\tabcolsep}{-3pt}
\scalebox{1.0}{
\begin{tabular} {|c|c|c|c|c|c|c|}\hline 
\multicolumn{4}{|c|}{\footnotesize Classifier type} & {\footnotesize CIFAR-10} & {\footnotesize CIFAR-100} & {\footnotesize ImageNet} \\ \hline
\multicolumn{2}{|c|}{\multirow{3}*{\footnotesize Target classifier}} & \multicolumn{2}{c|}{\footnotesize ResNet20} & {\footnotesize 0.91} & {\footnotesize --} & {\footnotesize --} \\ \cline{3-7}
									\multicolumn{2}{|c|}{} & \multicolumn{2}{c|}{\footnotesize WRN-22-4} & {\footnotesize --} & {\footnotesize 0.75} & {\footnotesize --} \\ \cline{3-7}
									\multicolumn{2}{|c|}{} & \multicolumn{2}{c|}{\footnotesize ResNet50} & {\footnotesize --} & {\footnotesize --} & {\footnotesize 0.75} \\ \hline

\multicolumn{2}{|c|}{\multirow{16}*{\makecell{\footnotesize Positive \\\\\footnotesize suspect \\\\\footnotesize classifiers}}}
						    	&	\multicolumn{2}{c|}{\footnotesize FTLL} & {\footnotesize 0.92} & {\footnotesize 0.75} & {\footnotesize 0.75} \\ \cline{3-7}
						\multicolumn{2}{|c|}{}  & \multicolumn{2}{c|}{\footnotesize FTAL} & {\footnotesize 0.92} & {\footnotesize 0.75} & {\footnotesize 0.76} \\ \cline{3-7}
						\multicolumn{2}{|c|}{}  & \multicolumn{2}{c|}{\footnotesize RTLL} & {\footnotesize 0.92} & {\footnotesize 0.75} & {\footnotesize 0.72} \\ \cline{3-7}
						\multicolumn{2}{|c|}{}  & \multicolumn{2}{c|}	{\footnotesize RTAL} & {\footnotesize 0.92} & {\footnotesize 0.75} & {\footnotesize 0.72} \\ \cline{3-7}
						\multicolumn{2}{|c|}{}  &	 \multirow{5}*{\footnotesize WP} & {\footnotesize p=0.1} & {\footnotesize 0.91} & {\footnotesize 0.74} & {\footnotesize 0.75}\\ \cline{4-7}
						\multicolumn{2}{|c|}{}  &	 {} 						  & {\footnotesize p=0.2} & {\footnotesize 0.91} & {\footnotesize 0.75} & {\footnotesize 0.75}\\ \cline{4-7}
						\multicolumn{2}{|c|}{}  &	 {} 						  & {\footnotesize p=0.3} & {\footnotesize 0.90} & {\footnotesize 0.74} & {\footnotesize 0.75}\\ \cline{4-7}
						\multicolumn{2}{|c|}{}  &	 {} 						  & {\footnotesize p=0.4} & {\footnotesize 0.90} & {\footnotesize 0.74} & {\footnotesize 0.73}\\ \cline{4-7}
						\multicolumn{2}{|c|}{}  &	 {} 						  & {\footnotesize p=0.5} & {\footnotesize -- }        & {\footnotesize 0.73} & {\footnotesize 0.72}\\ \cline{3-7}
						
						\multicolumn{2}{|c|}{}  &	 \multirow{7}*{\footnotesize FP} & {\footnotesize c=1/16} & {\footnotesize 0.91} & {\footnotesize 0.75} & {\footnotesize 0.73}\\ \cline{4-7}
						\multicolumn{2}{|c|}{}  &	 {}                                  & {\footnotesize c=2/16} & {\footnotesize 0.91} & {\footnotesize 0.74} & {\footnotesize 0.73}\\ \cline{4-7}
						\multicolumn{2}{|c|}{}  &	 {}                                  & {\footnotesize c=3/16} & {\footnotesize 0.91} & {\footnotesize 0.74} & {\footnotesize 0.72}\\ \cline{4-7}
						\multicolumn{2}{|c|}{}  &	 {}                                  & {\footnotesize c=4/16} & {\footnotesize 0.90} & {\footnotesize 0.73} & {\footnotesize --}\\ \cline{4-7}
						\multicolumn{2}{|c|}{}  &	 {}                                  & {\footnotesize c=5/16} & {\footnotesize 0.89} & {\footnotesize 0.73} & {\footnotesize --}\\ \cline{4-7}
						\multicolumn{2}{|c|}{}  &	 {}                                  & {\footnotesize c=6/16} & {\footnotesize 0.89} & {\footnotesize --}         & {\footnotesize --}\\ \cline{4-7}
						\multicolumn{2}{|c|}{}  &	 {}                                  & {\footnotesize c=7/16} & {\footnotesize 0.88} & {\footnotesize --}         & {\footnotesize --}\\ \hline
						
\multirow{15}*{\makecell {\footnotesize Negative \\\\ \footnotesize suspect \\\\ \footnotesize classifiers}} 
						& {\makecell{\footnotesize Same \\\footnotesize architecture}}
							& \multicolumn{2}{c|}{\footnotesize ResNet20}  & {\footnotesize [0.91,0.92]} & {\footnotesize --} & {\footnotesize --}\\ \cline{3-7}
					       ~ &    {\footnotesize DNNs}  & \multicolumn{2}{c|}{\footnotesize WRN-22-4} & {\footnotesize --} & {\footnotesize [0.74,0.76]} & {\footnotesize --}\\ \cline{2-7}
					       ~ &  \multirow{11}*{\makecell{\footnotesize Different \\\\ \footnotesize architecture \\\\ \footnotesize DNNs}} 
							 & \multicolumn{2}{c|}{\footnotesize LeNet-5} & {\footnotesize [0.64,0.67]} & {\footnotesize [0.31,0.34]} & {\footnotesize --}\\ \cline{3-7}
					       ~ &    ~ & \multicolumn{2}{c|}{\footnotesize VGG16} & {\footnotesize [0.93,0.94]} & {\footnotesize [0.68,0.70]} & {\footnotesize 0.71}\\ \cline{3-7}
					       ~ &    ~ & \multicolumn{2}{c|}{\footnotesize ResNet152} & {\footnotesize --} & {\footnotesize --} & {\footnotesize 0.77}\\ \cline{3-7}
					       ~ &    ~ & \multicolumn{2}{c|}{\footnotesize ResNet152V2} & {\footnotesize --} & {\footnotesize --} & {\footnotesize 0.78}\\ \cline{3-7}
					       ~ &    ~ & \multicolumn{2}{c|}{\footnotesize InceptionV3} & {\footnotesize --} & {\footnotesize --} & {\footnotesize 0.78}\\ \cline{3-7}
					       ~ &    ~ & \multicolumn{2}{c|}{\footnotesize InceptionResNetV2} & {\footnotesize --} & {\footnotesize --} & {\footnotesize 0.80}\\ \cline{3-7}
					       ~ &    ~ & \multicolumn{2}{c|}{\footnotesize Xception} & {\footnotesize --} & {\footnotesize --} & {\footnotesize 0.79}\\ \cline{3-7}				     
					       ~ &    ~ & \multicolumn{2}{c|}{\footnotesize MobileNet\tiny($\alpha$=1.0)} & {\footnotesize --} & {\footnotesize --} & {\footnotesize 0.70}\\ \cline{3-7}
					       ~ &    ~ & \multicolumn{2}{c|}{\footnotesize MobileNetV2\tiny($\alpha$=1.4)} & {\footnotesize --} & {\footnotesize --} & {\footnotesize 0.75}\\ \cline{3-7}
					       ~ &    ~ & \multicolumn{2}{c|}{\footnotesize DenseNet201} & {\footnotesize --} & {\footnotesize --} & {\footnotesize 0.77}\\ \cline{3-7}
					       ~ &    ~ & \multicolumn{2}{c|}{\footnotesize NASNetLarge} & {\footnotesize --} & {\footnotesize --} & {\footnotesize 0.83}\\  \cline{2-7}
					       ~ & \makecell{\footnotesize Random \\\footnotesize forests} & \multicolumn{2}{c|}{\footnotesize RF} 
  						      & {\footnotesize [0.40,0.41]} & {\footnotesize [0.15,0.16]} & {\footnotesize --}\\ \hline

\end{tabular}
}
\label{tab:accuracies}
\end{table}

\myparatight{Same-architecture neural network classifiers} It is well known that a neural network's loss function is non-convex. Therefore, different initializations may lead to different model parameters, which are different local minima of the loss function. Therefore, if two owners train their own model parameters using the same (publicly available) training data and neural network architecture, the two models should be treated as different ones. In other words, the fingerprint extracted for one model should not be verified for the other model. 

To test our fingerprinting method in such scenarios, we train 50 neural network classifiers that have the same architecture as the target classifier for CIFAR-10 or CIFAR-100 as suspect classifiers.  
Due to our limited computation resource, we do not train the same-architecture neural network classifiers for ImageNet.

\myparatight{Different-architecture neural network classifiers} For CIFAR-10 and CIFAR-100 datasets, we consider LeNet-5 ~\cite{lecun1998gradient} and the popular VGG16 architecture~\cite{simonyan2014very} as our different-architecture neural networks. Specifically, we train 10 models for each architecture using different initializations for both CIFAR-10 and CIFAR-100 datasets. 

For the ImageNet dataset, due to our limited computation resource, we leverage popular pre-trained models that are available in Keras~\cite{chollet2015keras}, instead of training our own models. For each family of classifiers, we select the one with the highest top-1 accuracy. For example, the pre-trained ResNet101 and ResNet152 achieve 76.42\% and 76.60\% top-1 accuracy, respectively, as reported by Keras documentation. Therefore, we choose ResNet152 as the representative for the ResNet family.  Overall, we consider Xception~\cite{chollet2017xception}, VGG16~\cite{simonyan2014very}, ResNet152~\cite{he2016deep}, ResNet152V2~\cite{he2016identity}, ResNeXt101~\cite{xie2017aggregated}, InceptionV3~\cite{szegedy2016rethinking}, InceptionResNetV2~\cite{szegedy2017inception}, MobileNet($\alpha$=1.0)~\cite{howard2017mobilenets}, MobileNetV2($\alpha$=1.4)~\cite{sandler2018mobilenetv2}, DenseNet201~\cite{huang2017densely}, and NASNetLarge~\cite{zoph2018learning}. Note that the input size for different models might be different. We resize the data points in our fingerprint to fit the input size of different architectures.

\myparatight{Random forest (RF)} Apart from neural networks, we also consider random forest as suspect classifiers. Specifically, for each of the three datasets, we use scikit-learn with the default settings to train 10 random forests, each of which has 20 trees. We note that random forests have relatively low classification accuracies. However, our goal is not to design accurate random forest classifiers. Instead, our goal is to show that our fingerprinting method can distinguish between neural networks and random forests. Note that  we did not train random forest classifiers for the ImageNet dataset because  our memory is not large enough. 

\myparatight{Positive vs. negative suspect classifiers} Our goal is that the post-processed versions of the target classifier are verified, while the other suspect classifiers are not verified. For convenience, we call the post-processed versions of the target classifier \emph{positive} suspect classifiers and the remaining ones \emph{negative} suspect classifiers. Table~\ref{tab:datasets} summarizes the number of positive and negative suspect classifiers for each dataset. 

Table ~\ref{tab:accuracies} shows the testing accuracies of the target and suspect classifiers.

\subsubsection{Compared Methods} \label{sec:imp_details} 
A key component of our IPGuard is to find data points  near a target classifier's classification boundary. An adversarial example method could be used to find such data points. Therefore, we compare with existing adversarial example methods including FGSM~\cite{goodfellow2014explaining}, IGSM~\cite{kurakin2016adversarial}, and CW-$L_2$~\cite{carlini2017towards}. 

\myparatight{Random} This method randomly samples a data point from the data domain as a fingerprinting data point. 

\myparatight{FGSM~\cite{goodfellow2014explaining} and IGSM~\cite{kurakin2016adversarial}} Both FGSM and IGSM have a parameter $\epsilon$, which bounds the noise in an adversarial example. We will explore different $\epsilon$. IGSM further has a step size $\alpha$. Since we use image datasets, we set $\alpha=\frac{1}{255}$, which is the minimal step size. Moreover, we set the maximum number of iterations of IGSM to be 1,000. 

\myparatight{CW-$L_2$~\cite{carlini2017towards}}  CW-$L_2$ has a parameter $k$, which controls the confidence of an adversarial example. We will explore the impact of $k$ on the performance of CW-$L_2$.  CW-$L_2$ leverages an Adam optimizer and we set the learning rate to be 0.001 on CIFAR-10 and CIFAR-100, and 0.1 on ImageNet. We use the implementation of CW-$L_2$ published by its authors. 

\begin{table}[!t]\renewcommand{\arraystretch}{1.0}
\centering
\caption{Suffix that represents a combination of initializing a data point and selecting target label. The rows represent two ways of initializations, and the columns represent two ways of selecting the target label.}
\centering
\begin{tabular} {|c|c|c|}\hline 
{\small } & {\small Random label} & {\small Least-likely label}\\ \hline
{\small Training example} & {\small -TR} & {\small -TL}\\ \hline
{\small  Random example} & {\small -RR} & {\small -RL}\\ \hline
\end{tabular}
\label{tab:postfix}
\end{table}
 
\myparatight{Our IPGuard} Our IPGuard also has a parameter $k$, which balances between robustness and uniqueness. Like CW-$L_2$, IPGuard also leverages Adam optimizer and we use the same learning rates as CW-$L_2$. IPGuard runs the Adam optimizer until we find a data point that makes the value of the objective function  0 or we have reached the maximum number of iterations. We set the maximum number of iterations to be 1,000. In our experiments, the iterative process always stops  before 1,000 iterations. In fact, IPGuard only needs several iterations in finding most fingerprinting data points.

\subsubsection{Initialization and target labels} Our fingerprint consists of $n$ data points near the target classifier's classification boundary with their labels predicted by the target classifier. In our experiments, we set $n=100$. We also tried larger $n$, e.g., $n=10,000$. However, IPGuard and existing efficient adversarial example methods such as FGSM and IGSM have negligible performance improvement (measured by ARUC that we will introduce in the next part). To find one such data point $x$, we need to initialize a data point and pick a target label, and then we use a compared method to find $x$. We consider two ways of initialization and two ways of selecting the target label. 

\myparatight{Initialization} We consider the following two ways to initialize a data point. 

\begin{itemize}
\item {\bf Training example (T).} In this method, we randomly sample a training example as the  initialized data point. 
\item {\bf Random (R).} In this method, we randomly sample a data point from the entire data domain as the initialized data point. Since we are using image datasets, we uniformly sample from $[0,1]^d$, where $d$ is the dimension of the images. 
\end{itemize}

\begin{figure*}[!t]
\centering
{\includegraphics[width=0.98 \textwidth]{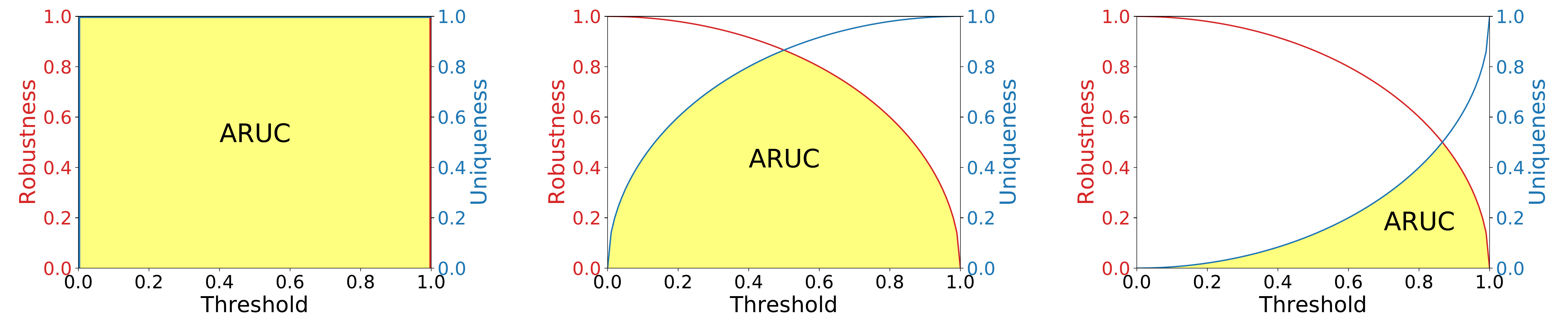}}
\caption{Illustration of robustness-uniqueness curves and our metric ARUC to jointly measure robustness and uniqueness. \emph{Left:} A perfect ARUC (i.e., AURC=1), where both robustness and uniqueness are 1 for any matching rate threshold. \emph{Middle:} A mediocre ARUC, where both robustness and uniqueness are large only when the matching rate threshold is around 0.5. \emph{Right:} A bad ARUC, where no matching rate threshold makes both robustness and uniqueness large.}
\label{ARUC}
\end{figure*}

\myparatight{Target label selection}  We consider the following two ways to select a target label. 
\begin{itemize}
\item {\bf Random (R).} In this method, we randomly select a  label as the target label for a data point. 
\item {\bf Least-likely label (L).} In this method, we select the least-likely label of an initialized data point as the target label. The least-likely label of a data point is the label that has the smallest logit predicted by the target classifier. In other words, the target classifier is least confident at predicting the least-likely label as the label for the initialized data point.  We speculate that the target classifier's classification boundary near such target label and initialized data point is more unique. 
\end{itemize}

We will compare the two initialization methods and the two methods of selecting target label. For convenience, we use a suffix ``-TR'', ``-TL'', ``-RR'', or ``-RL'' to represent a combination of initialization and  target label selection. Table \ref{tab:postfix} summarizes the suffixes. For instance, FGSM-TL means that we initialize a data point as a training example, select the least-likely label as the target label, and use FGSM to find a targeted adversarial example as a fingerprinting data point.  

\subsubsection{Evaluation Metrics} As described in Section~\ref{property}, we desire 5  properties for a fingerprinting method. All compared  methods have no accuracy loss of the target classifier. Moreover, all compared  methods achieve the effectiveness property. Therefore, we compare IPGuard and existing adversarial example methods with respect to robustness, uniqueness, and efficiency properties. 

\myparatight{Matching rate} For a suspect classifier, we query the classifier's API for the labels of the fingerprinting data points. \emph{Matching rate} is the fraction of fingerprinting data points whose labels predicted by the suspect classifier match those predicted by the target classifier. A positive suspect classifier (i.e., a post-processed version of the target classifier) should have a large  matching rate, while a negative suspect classifier should have a low matching rate. 

\begin{figure*}[!t]
\centering
\subfloat[CIFAR-10]{\includegraphics[width=0.3\textwidth]{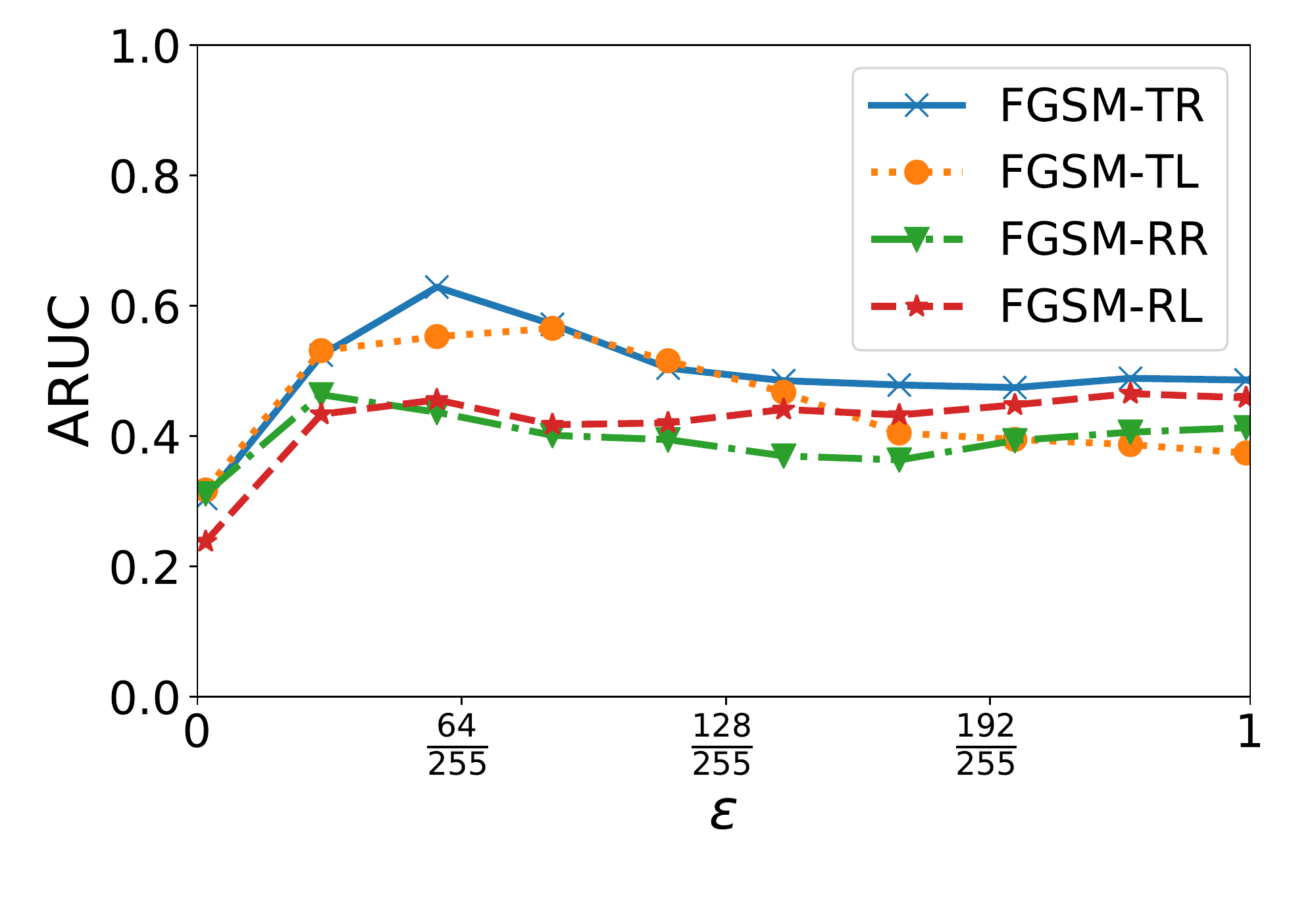}}
\subfloat[CIFAR-100]{\includegraphics[width=0.3 \textwidth]{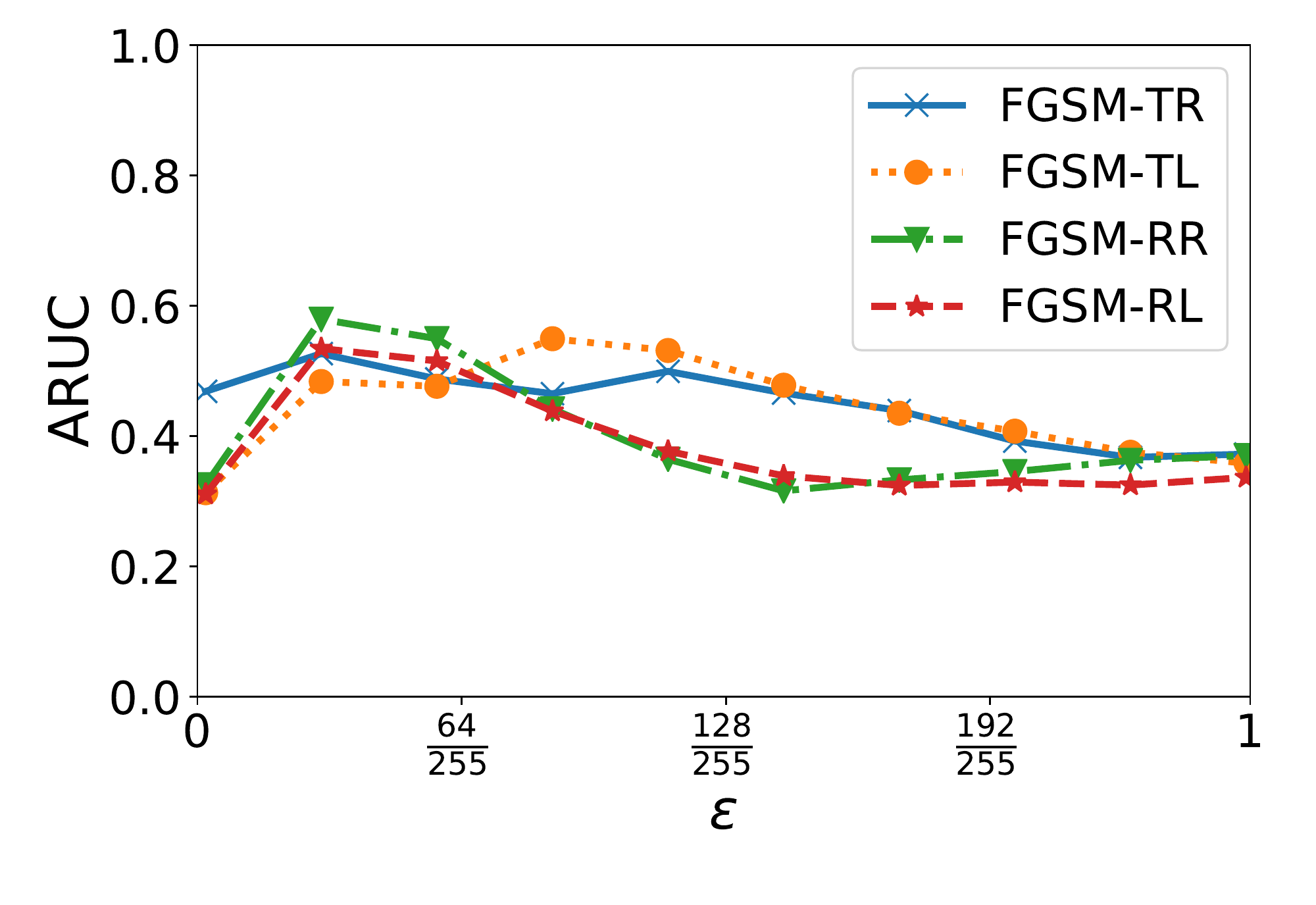}}
\subfloat[ImageNet]{\includegraphics[width=0.3 \textwidth]{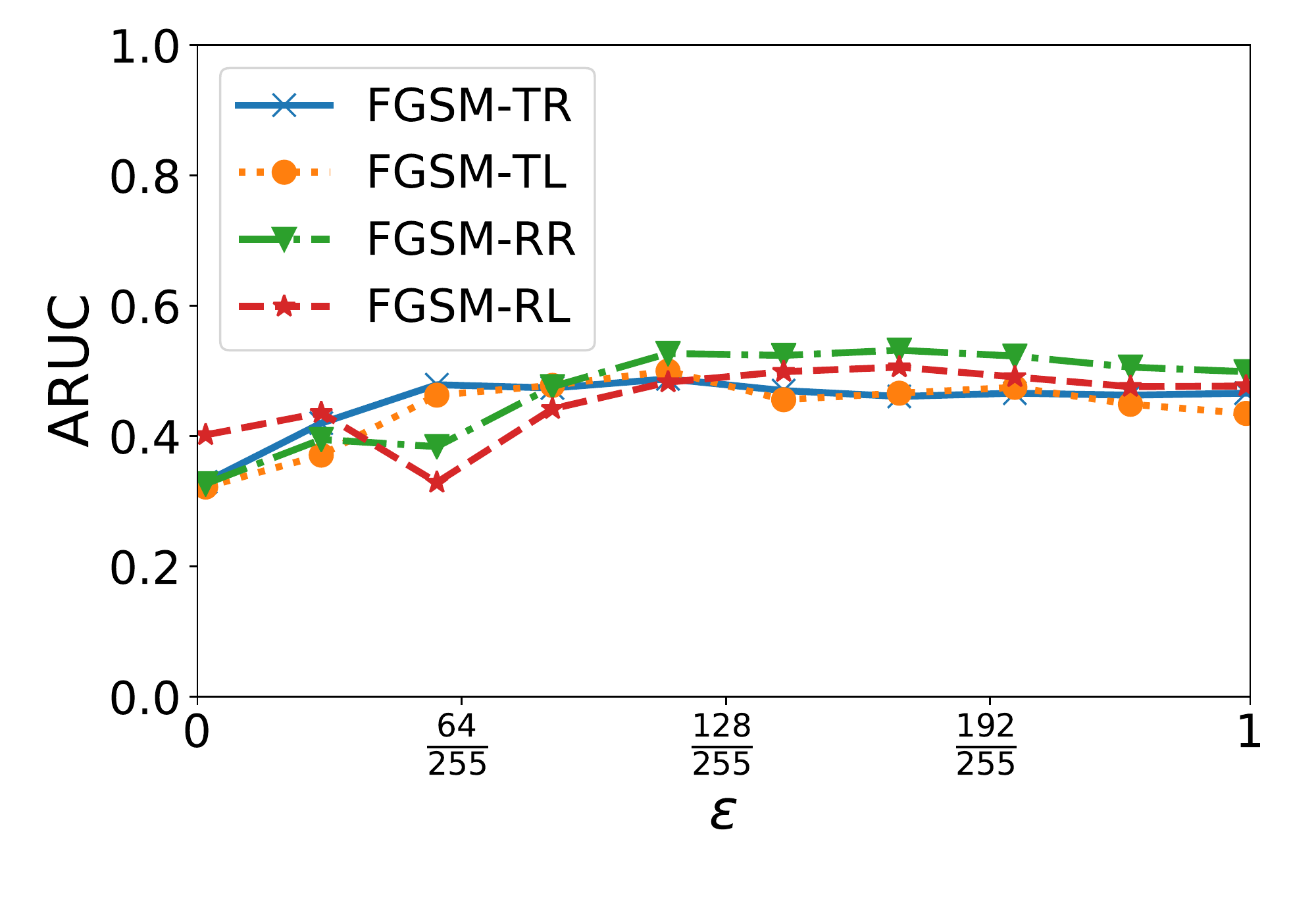}}
\vspace{-2mm}
\caption{Impact of $\epsilon$ on ARUC for FGSM.}
\label{fgsm}
\end{figure*}

\begin{figure*}[!t]
\vspace{-4mm}
\centering
\subfloat[CIFAR-10]{\includegraphics[width=0.3 \textwidth]{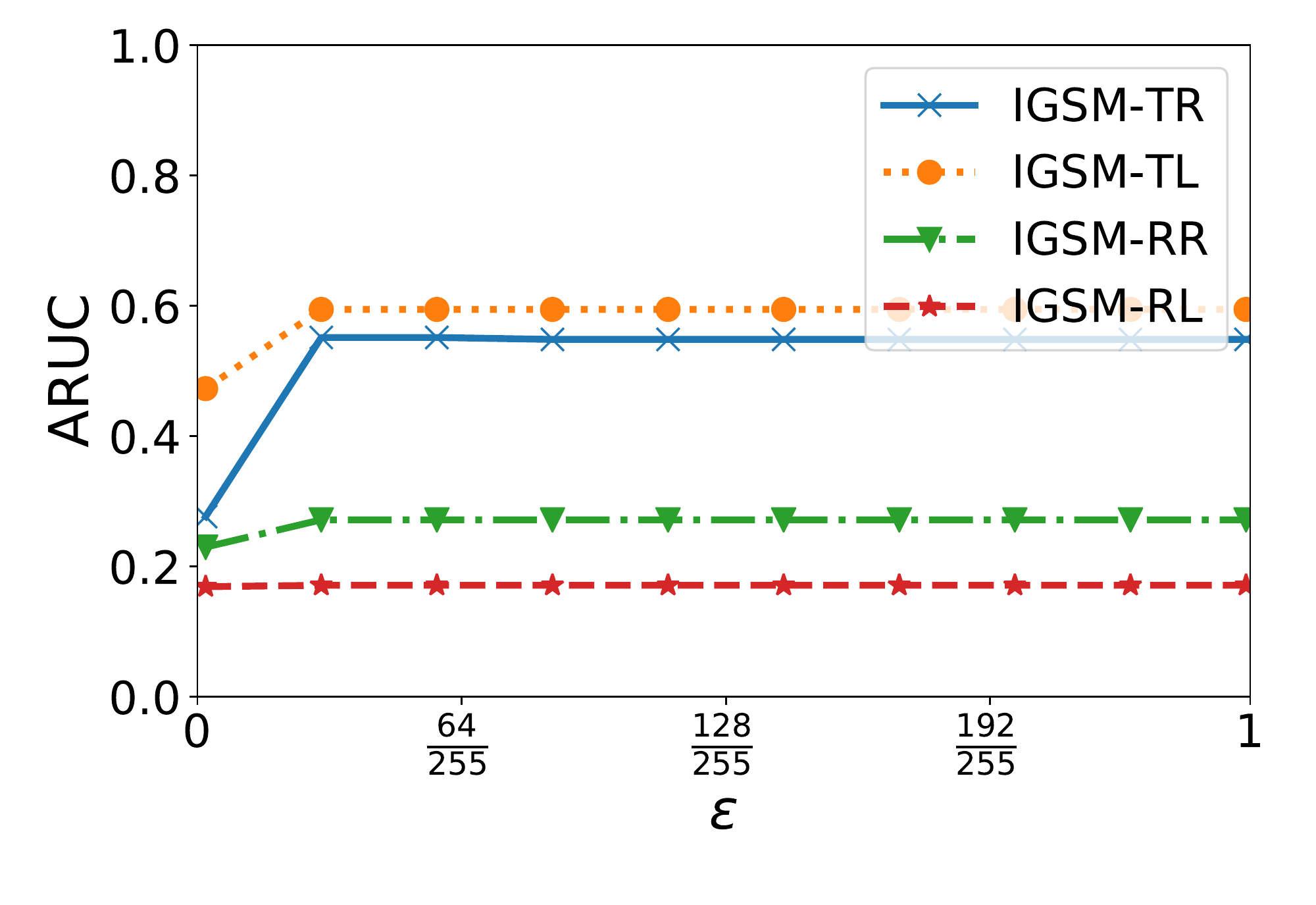}}
\subfloat[CIFAR-100]{\includegraphics[width=0.3 \textwidth]{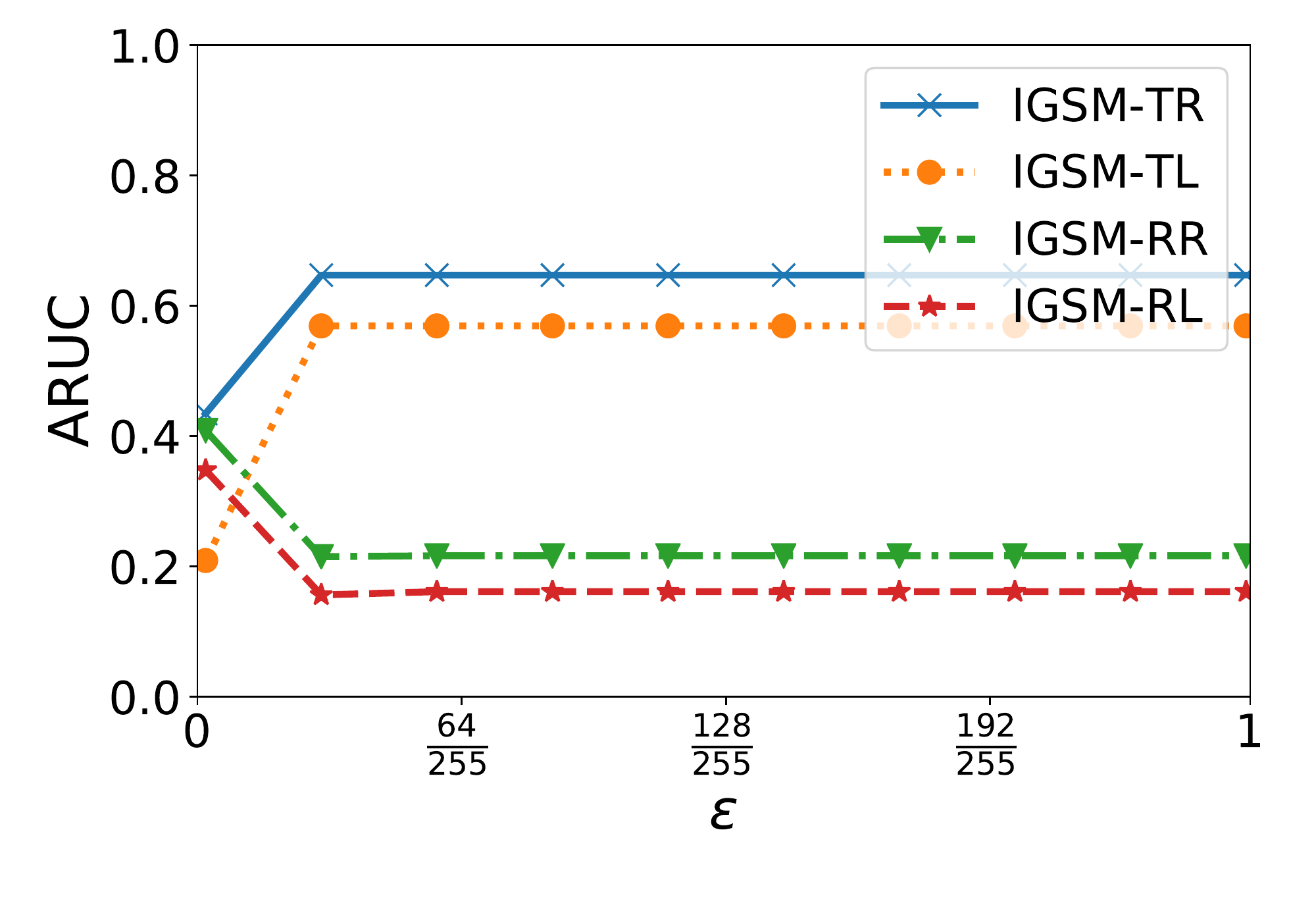}}
\subfloat[ImageNet]{\includegraphics[width=0.3 \textwidth]{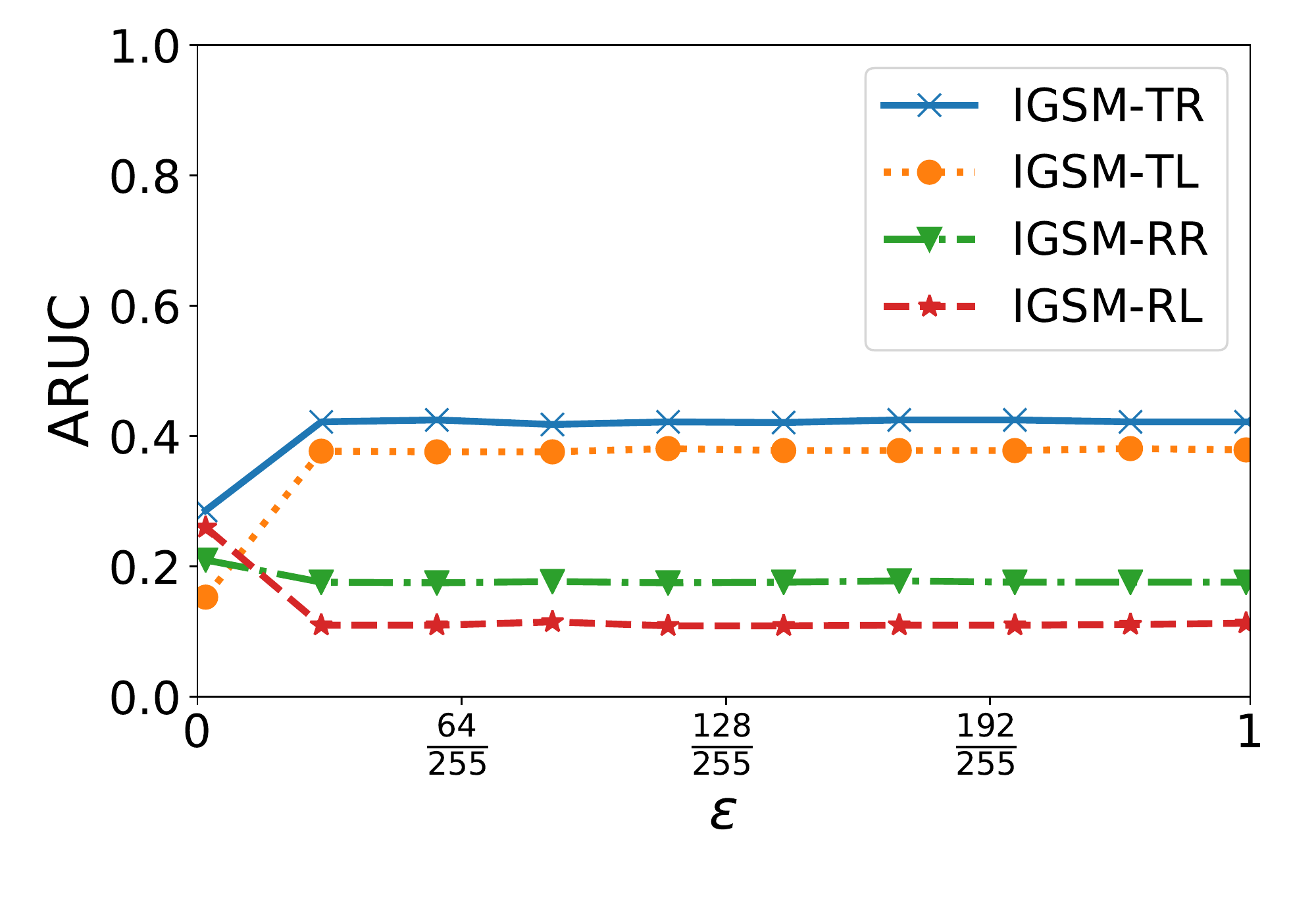}}
\vspace{-2mm}
\caption{Impact of $\epsilon$ on ARUC for IGSM.}
\label{igsm}
\end{figure*}

\myparatight{ARUC} Given a matching rate threshold, the Verify function produces 1 if a suspect classifier has a matching rate that is greater than or equal to the threshold. 
 We use the fraction of positive suspect classifiers that are verified to measure the \emph{robustness} of a  method, and we use the fraction of negative suspect classifiers that are not verified to measure the \emph{uniqueness} of a  method.  

\begin{figure*}[!t]
\vspace{-4mm}
\centering
\subfloat[CIFAR-10]{\includegraphics[width=0.3 \textwidth]{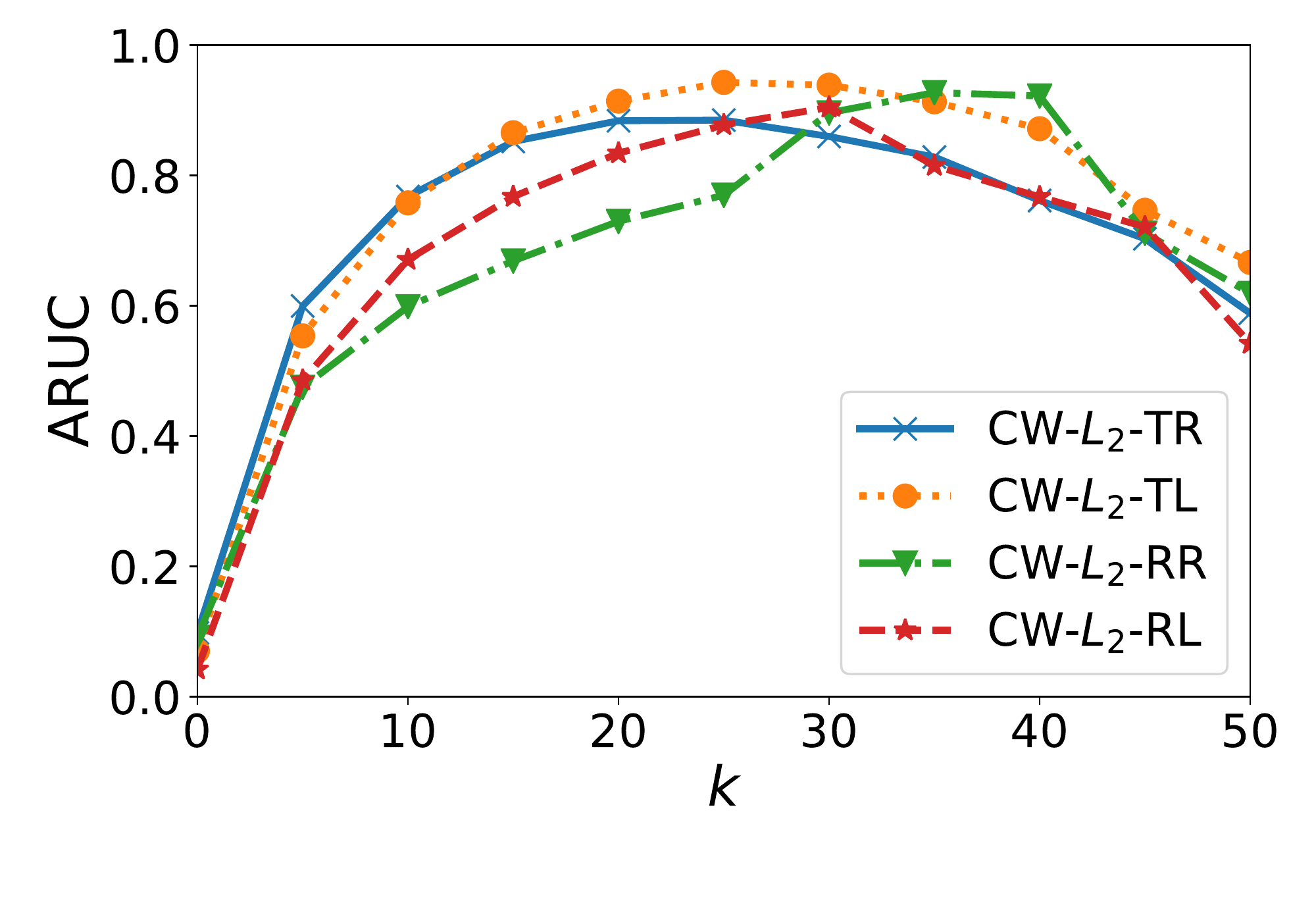}}
\subfloat[CIFAR-100]{\includegraphics[width=0.3 \textwidth]{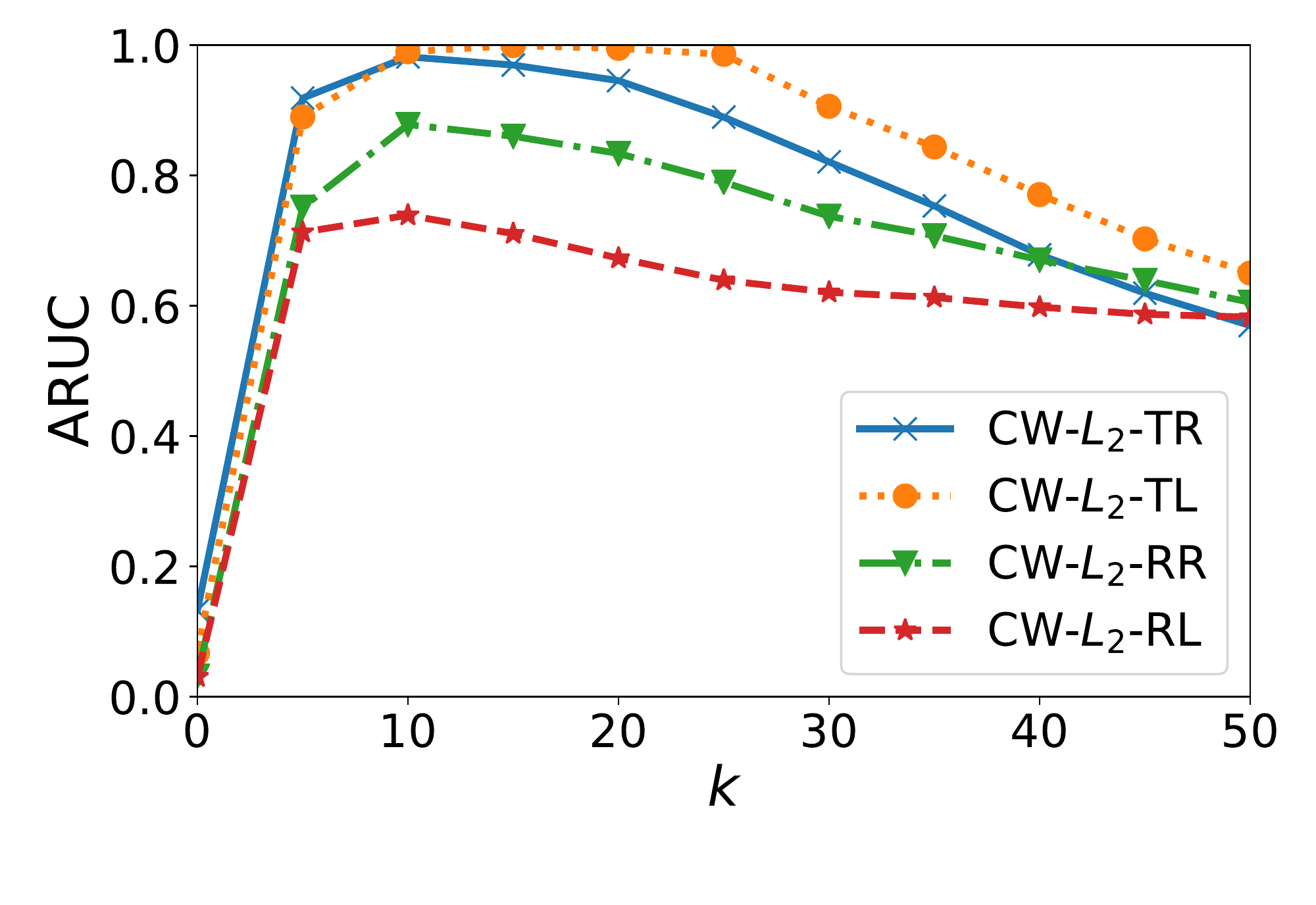}}
\subfloat[ImageNet]{\includegraphics[width=0.3 \textwidth]{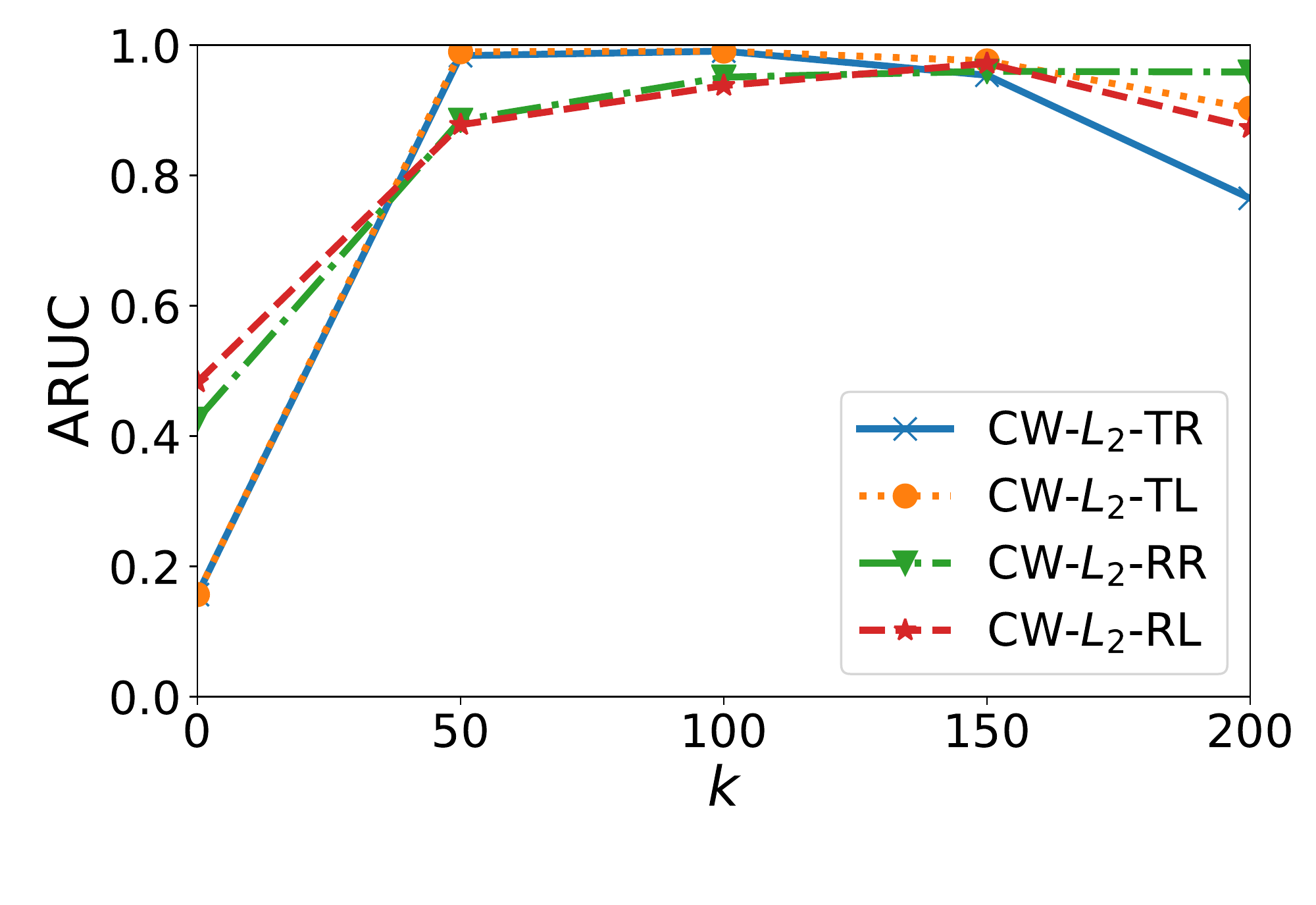}}
\vspace{-2mm}
\caption{Impact of $k$ on ARUC for CW-$L_2$.}
\label{cw}
\end{figure*}

\begin{figure*}[!t]
\vspace{-4mm}
\centering
\subfloat[CIFAR-10]{\includegraphics[width=0.3 \textwidth]{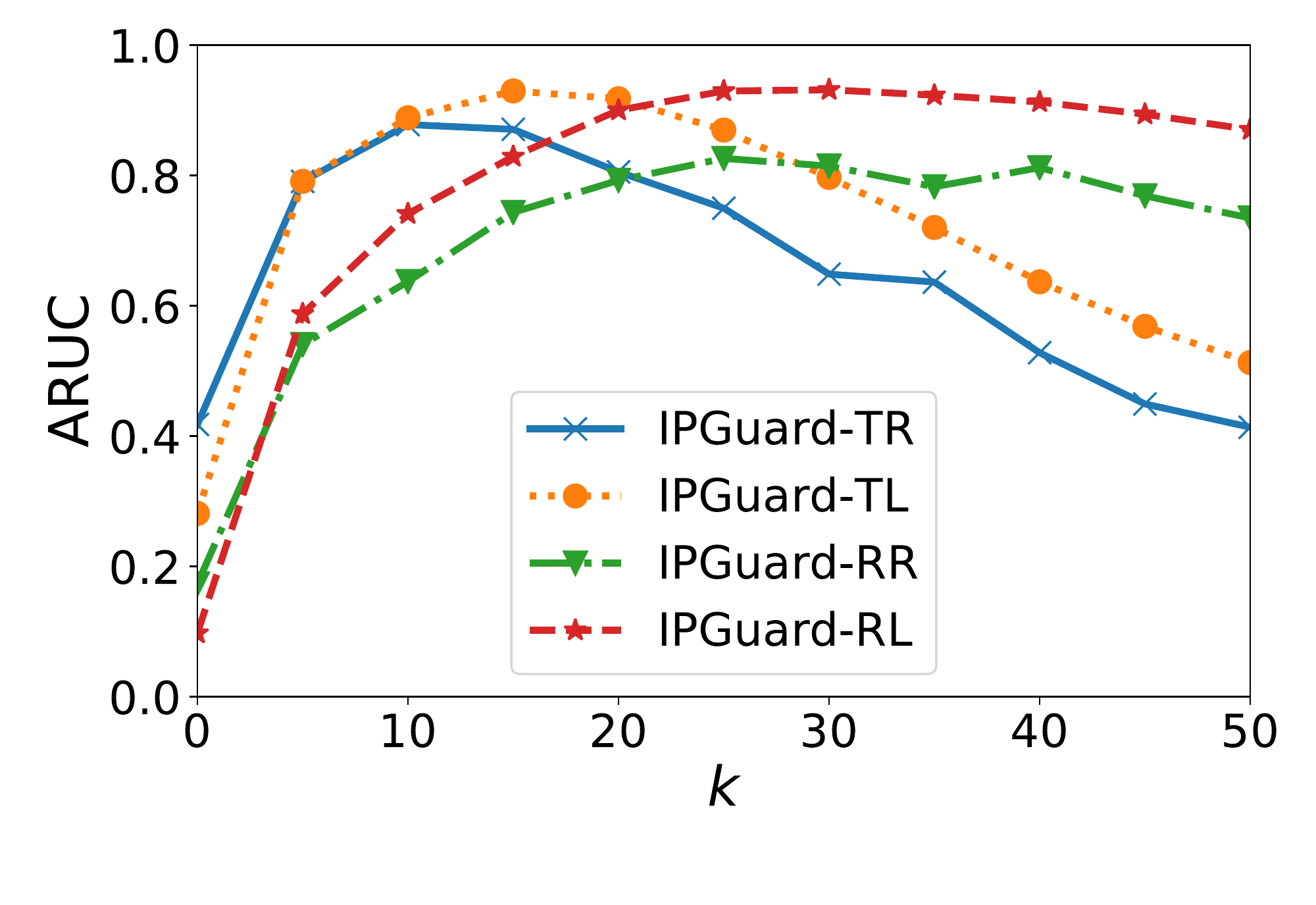}}
\subfloat[CIFAR-100]{\includegraphics[width=0.3 \textwidth]{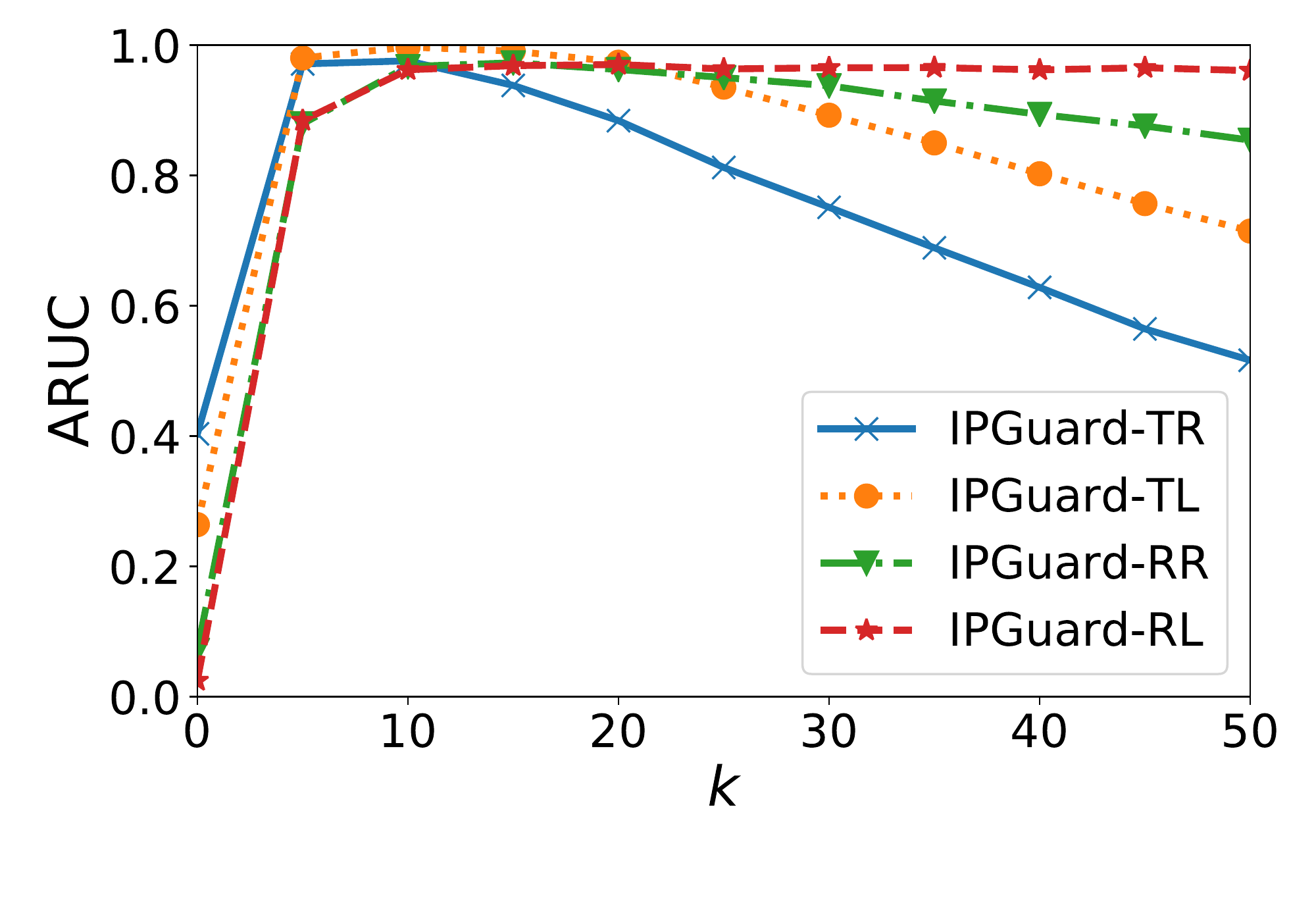}}
\subfloat[ImageNet]{\includegraphics[width=0.3 \textwidth]{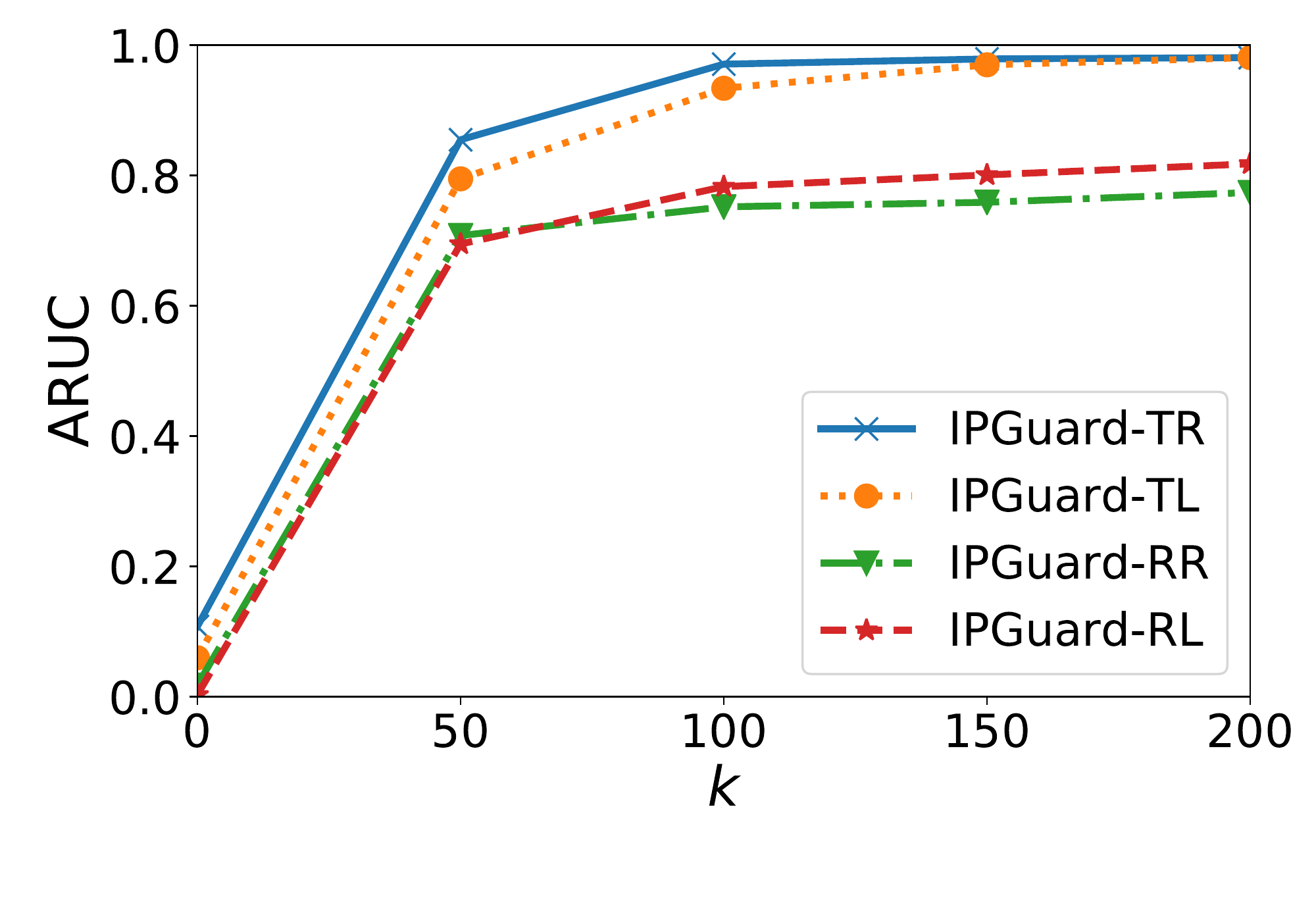}}
\vspace{-2mm}
\caption{Impact of $k$ on ARUC for IPGuard.}
\label{IPGuard}
\end{figure*}

 The matching rate threshold controls the robustness-uniqueness tradeoffs. 
 In particular, robustness decreases and uniqueness increases as the matching rate threshold increases. One way to avoid the impact of the selection of matching rate threshold and consistently compare different methods is to use AUC~\cite{bradley1997use}, a standard machine learning evaluation metric, to jointly measure  the robustness and uniqueness of a method. Specifically, we can rank the positive and negative suspect classifiers in a decreasing order according to their matching rates. Then, AUC is essentially the probability that a randomly sampled positive suspect classifier ranks higher than a randomly sampled negative suspect classifier. However, AUC is insufficient to jointly measure robustness and uniqueness in our problem. Specifically, AUC does not characterize the gaps between the matching rates of positive suspect classifiers and those of the negative suspect classifiers. For instance, once all positive suspect classifiers have larger matching rates than all negative suspect classifiers, AUC is 1 no matter how large the gap between the matching rates is. However, such gap is important in our problem. Specifically, the positive and negative suspect classifiers we evaluate may be a subset of suspect classifiers in practice. A larger gap between the matching rates of our evaluated positive suspect classifiers and those of our evaluated negative suspect classifiers means that it is more likely to select a matching rate threshold to  achieve both large robustness and uniqueness in practice.

Therefore, we propose the \emph{Area under the  Robustness-Uniqueness Curves (ARUC)} as a metric to jointly evaluate the robustness and uniqueness of a  method. Specifically, we draw the robustness and uniqueness curves in the same graph as we increase the matching rate threshold from 0 to 1. The two curves intersect at a certain point. We define ARUC as the area under the intersected Robustness-Uniqueness curves. Figure~\ref{ARUC} illustrates ARUC in three scenarios. ARUC ranges from 0 to 1. A larger ARUC indicates that it is more likely to have  large robustness and  uniqueness  at the same time, i.e., for a wider range of matching rate threshold, both robustness and  uniqueness are large. 

Formally,  ARUC is defined as follows:
\begin{equation}
ARUC = \int_{0}^{1} \min\{R(\tau), U(\tau)\}d\tau, \label{eq:calc_aruc}
\end{equation}
where $\tau$ is a matching rate threshold, and $R(\tau)$ and $U(\tau)$  are the robustness and uniqueness when the matching rate threshold is $\tau$, respectively. We evenly divide the range [0,1] to $r$ intervals and use the rightmost point in each interval to represent the interval. Then, we can approximate ARUC as follows:
\begin{equation}
ARUC = \frac{1}{r}\sum_{\tau'=1}^{r}{\min\{R(\frac{\tau'}{r}), U(\frac{\tau'}{r})\}}.
\end{equation}
In our experiments, we set $r=100$.

\myparatight{Efficiency} In practice, deep neural network classifiers often have large scales. Therefore, the fingerprinting method should be scalable. We use the running time of finding the  fingerprinting data points to measure the efficiency of a method.  

\subsection{Results}

\myparatight{ARUC}  Figure~\ref{fgsm} and Figure~\ref{igsm} respectively show the impact of $\epsilon$ on the ARUC for FGSM and IGSM, where the four combinations of initialization and target label selection are considered (please refer to Table~\ref{tab:postfix} for the definition of the suffixes).  For FGSM, ARUC first increases and then decreases or fluctuates as $\epsilon$ increases. For IGSM,  ARUC first increases and then keeps stable as $\epsilon$ increases. ARUC keeps stable for large $\epsilon$ because IGSM moves a step $\alpha$ in each iteration no matter how large $\epsilon$ is and stops as soon as it finds an adversarial example. We observe that no matter how we select $\epsilon$, FGSM and IGSM have ARUCs that are less than 0.7.  

Figure~\ref{cw} and Figure~\ref{IPGuard} respectively show the impact of $k$ on ARUC for CW-$L_2$ and IPGuard, where $k$ controls the confidence of adversarial examples for CW-$L_2$ and the distance from the fingerprinting data points to the classification boundary for IPGuard. For both  methods, ARUC first increases and then decreases as $k$ increases (except that IPGuard has not started to decrease before $k=200$ on ImageNet). This is because, as $k$ increases, the fingerprinting data points are further away from the target classifier's classification boundary. Therefore, the matching rates of both positive and negative suspect classifiers increase, which means that robustness increases and uniqueness decreases. Moreover, matching rates of positive suspect classifiers increase faster than those of negative suspect classifiers when $k$ is small, but slower than those of negative suspect classifiers when $k$ is large. 
 As a result,  ARUC first increases and then decreases. 
For both CW-$L_2$ and IPGuard, the combination ``-TL'' can achieve the best ARUCs.  Moreover, for IPGuard,  least-likely label outperforms random label when $k$ is large no matter how we initialize the fingerprinting data points, i.e., IPGuard-TL outperforms  IPGuard-TR and IPGuard-RL outperforms  IPGuard-RR, except that IPGuard-TL and IPGuard-TR achieve comparable ARUCs on ImageNet when $k$ is large.  We speculate the reason is that the classification boundary near the initialized data points and their least-likely labels is more unique. 

Table~\ref{tab:ARUC} shows the best ARUC that each compared method can achieve via proper parameter setting for each dataset. As we can see, CW-$L_2$ and IPGuard achieve comparable ARUCs; and they substantially outperform other methods. 

\begin{table}[!t]\renewcommand{\arraystretch}{1.00}
\centering
\caption{ARUCs of the compared methods. }
\centering
\scalebox{1.0}{
\begin{tabular} {|c|c|c|c|c|}\hline 
{\small } & {\small CIFAR-10} & {\small CIFAR-100} & {\small ImageNet}\\ \hline
{\small Random} & {\small 0.35} & {\small 0.52} & {\small 0.22}\\ \hline
{\small FGSM} & {\small 0.63} & {\small 0.58} & {\small 0.53}\\ \hline
{\small IGSM} & {\small 0.59} & {\small 0.65} & {\small 0.43}\\ \hline
{\small CW-$L_2$} & {\small{0.95}} & {\small 1.00} & {\small 0.99}\\ \hline
{\small IPGuard} & {\small 0.94} & {\small 1.00} & {\small 0.98}\\ \hline
\end{tabular}
}
\label{tab:ARUC}
\end{table}

\begin{table}[!t]\renewcommand{\arraystretch}{1.0}
\centering
\caption{Matching rates of the compared methods for positive and negative suspect classifiers on CIFAR-10.}
\centering
\addtolength{\tabcolsep}{-3.2pt}
\scalebox{1.0}{
\begin{tabular} {|c|c|c|c|c|c|c|c|}\hline 
\multicolumn{4}{|c|}{\footnotesize Suspect classifier} & {\footnotesize FGSM} & {\footnotesize IGSM}  & {\footnotesize CW-$L_2$} & {\footnotesize IPGuard} \\ \hline

\multicolumn{2}{|c|}{\multirow{15}*{\makecell{\footnotesize Positive \\\\\footnotesize suspect \\\\\footnotesize classifiers}}}
						    	&	\multicolumn{2}{c|}{\footnotesize FTLL} & {\footnotesize 0.90} & {\footnotesize 0.99}  & {\footnotesize 1.00} & {\footnotesize 1.00} \\ \cline{3-8}
						\multicolumn{2}{|c|}{}  & \multicolumn{2}{c|}{\footnotesize FTAL} & {\footnotesize 0.92} & {\footnotesize 0.90} & {\footnotesize 1.00} & {\footnotesize 1.00}\\ \cline{3-8}
						\multicolumn{2}{|c|}{}  & \multicolumn{2}{c|}{\footnotesize RTLL} & {\footnotesize 0.87} & {\footnotesize 0.99} & {\footnotesize 1.00} & {\footnotesize 1.00}\\ \cline{3-8}
						\multicolumn{2}{|c|}{}  &	 \multicolumn{2}{c|}{\footnotesize RTAL} & {\footnotesize 0.90} & {\footnotesize 0.66} & {\footnotesize 1.00} & {\footnotesize 1.00}\\ \cline{3-8}
						
						\multicolumn{2}{|c|}{}  &	 \multirow{4}*{\footnotesize WP} & {\footnotesize p=0.1} & {\footnotesize 0.86} & {\footnotesize 0.91} & {\footnotesize 1.00} & {\footnotesize 1.00} \\ \cline{4-8}
						\multicolumn{2}{|c|}{}  &	 {}						 & {\footnotesize p=0.2} & {\footnotesize 0.86} & {\footnotesize 0.91} & {\footnotesize 1.00} & {\footnotesize 1.00} \\ \cline{4-8}
						\multicolumn{2}{|c|}{}  &	 {}						 & {\footnotesize p=0.3} & {\footnotesize 0.89} & {\footnotesize 0.88} & {\footnotesize 1.00} & {\footnotesize 1.00} \\ \cline{4-8}
						\multicolumn{2}{|c|}{}  &	 {}						 & {\footnotesize p=0.4} & {\footnotesize 0.90} & {\footnotesize 0.70} & {\footnotesize 1.00} & {\footnotesize 1.00} \\ \cline{3-8}
												
						\multicolumn{2}{|c|}{}  &	 \multirow{7}*{\footnotesize FP} & {\footnotesize c=1/16} & {\footnotesize 0.78} & {\footnotesize 0.67} & {\footnotesize 1.00} & {\footnotesize 1.00} \\ \cline{4-8}
						\multicolumn{2}{|c|}{}  &	 {}						 & {\footnotesize c=2/16} & {\footnotesize 0.80} & {\footnotesize 0.41} & {\footnotesize 1.00} & {\footnotesize 1.00} \\ \cline{4-8}
						\multicolumn{2}{|c|}{}  &	 {}						 & {\footnotesize c=3/16} & {\footnotesize 0.87} & {\footnotesize 0.37} & {\footnotesize 1.00} & {\footnotesize 1.00} \\ \cline{4-8}
						\multicolumn{2}{|c|}{}  &	 {}						 & {\footnotesize c=4/16} & {\footnotesize 0.82} & {\footnotesize 0.18} & {\footnotesize 0.99} & {\footnotesize 0.99} \\ \cline{4-8}
						\multicolumn{2}{|c|}{}  &	 {}						 & {\footnotesize c=5/16} & {\footnotesize 0.79} & {\footnotesize 0.17} & {\footnotesize 0.97} & {\footnotesize 0.94} \\ \cline{4-8}
						\multicolumn{2}{|c|}{}  &	 {}						 & {\footnotesize c=6/16} & {\footnotesize 0.74} & {\footnotesize 0.14} & {\footnotesize 0.89} & {\footnotesize 0.85} \\ \cline{4-8}
						\multicolumn{2}{|c|}{}  &	 {}						 & {\footnotesize c=7/16} & {\footnotesize 0.77} & {\footnotesize 0.09} & {\footnotesize 0.70} & {\footnotesize 0.71} \\ \hline

\multirow{6}*{\makecell {\footnotesize Negative \\ \footnotesize suspect \\ \footnotesize classifiers}} 
						& {\makecell{\footnotesize Same \\\footnotesize architecture \\\footnotesize DNNs}}
							&  \multicolumn{2}{c|}{\footnotesize ResNet20} & {\footnotesize [0.06,0.66]} & {\footnotesize [0.00,0.01]} & {\footnotesize [0.00,0.07]} & {\footnotesize [0.00,0.09]}\\\cline{2-8}
					       ~ &  {\makecell{\footnotesize Different \\\footnotesize architecture \\}} 
							 &  \multicolumn{2}{c|}{\footnotesize LeNet-5} & {\footnotesize [0.08,0.49]} & {\footnotesize [0.00,0.03]} & {\footnotesize [0.00,0.03]} & {\footnotesize [0.00,0.03]}\\ \cline{3-8}
					       ~ &    {\footnotesize DNNs} &  \multicolumn{2}{c|}{\footnotesize VGG16} & {\footnotesize [0.08,0.78]} & {\footnotesize [0.00,0.01]} & {\footnotesize [0.00,0.00]} & {\footnotesize [0.00,0.01]}\\ \cline{2-8}
					       ~ & \makecell{\footnotesize Random\\\footnotesize forests} & \multicolumn{2}{c|}{\footnotesize RF} 
  						      & {\footnotesize [0.02,0.09]} & {\footnotesize [0.00,0.01]} & {\footnotesize [0.00,0.00]} & {\footnotesize [0.00,0.00]}\\ \hline

\end{tabular}
}
\label{matchingrate-CIFAR10}
\end{table}

\begin{table}[!t]\renewcommand{\arraystretch}{1.0}
\centering
\caption{Matching rates of the compared methods for positive and negative suspect classifiers on CIFAR-100.}
\centering
\addtolength{\tabcolsep}{-3.2pt}
\scalebox{1.0}{
\begin{tabular} {|c|c|c|c|c|c|c|c|}\hline 
\multicolumn{4}{|c|}{\footnotesize Suspect classifier} & {\footnotesize FGSM} & {\footnotesize IGSM} & {\footnotesize CW-$L_2$} & {\footnotesize IPGuard} \\ \hline

\multicolumn{2}{|c|}{\multirow{14}*{\makecell{\footnotesize Positive \\\\\footnotesize suspect \\\\\footnotesize classifiers}}}
						    	&	\multicolumn{2}{c|}{\footnotesize FTLL} & {\footnotesize 0.99} & {\footnotesize 0.98} & {\footnotesize 1.00} & {\footnotesize 1.00} \\ \cline{3-8}
						\multicolumn{2}{|c|}{}  & \multicolumn{2}{c|}{\footnotesize FTAL} & {\footnotesize 0.86} & {\footnotesize 0.92} & {\footnotesize 1.00} & {\footnotesize 1.00}\\ \cline{3-8}
						\multicolumn{2}{|c|}{}  & \multicolumn{2}{c|}{\footnotesize RTLL} & {\footnotesize 0.92} & {\footnotesize 0.81} & {\footnotesize 1.00} & {\footnotesize 1.00}\\ \cline{3-8}
						\multicolumn{2}{|c|}{}  &	 \multicolumn{2}{c|}{\footnotesize RTAL} & {\footnotesize 0.76} & {\footnotesize 0.49} & {\footnotesize 1.00} & {\footnotesize 1.00}\\ \cline{3-8}
						
						\multicolumn{2}{|c|}{}  &	 \multirow{5}*{\footnotesize WP} & {\footnotesize p=0.1} & {\footnotesize 0.76} & {\footnotesize 0.88} & {\footnotesize 1.00} & {\footnotesize 1.00} \\ \cline{4-8}
						\multicolumn{2}{|c|}{}  &	 {}						 & {\footnotesize p=0.2} & {\footnotesize 0.72} & {\footnotesize 0.88} & {\footnotesize 1.00} & {\footnotesize 1.00} \\ \cline{4-8}
						\multicolumn{2}{|c|}{}  &	 {}						 & {\footnotesize p=0.3} & {\footnotesize 0.70} & {\footnotesize 0.83} & {\footnotesize 1.00} & {\footnotesize 1.00} \\ \cline{4-8}
						\multicolumn{2}{|c|}{}  &	 {}						 & {\footnotesize p=0.4} & {\footnotesize 0.54} & {\footnotesize 0.78} & {\footnotesize 1.00} & {\footnotesize 1.00} \\ \cline{4-8}
						\multicolumn{2}{|c|}{}  &	 {}						 & {\footnotesize p=0.5} & {\footnotesize 0.56} & {\footnotesize 0.66} & {\footnotesize 1.00} & {\footnotesize 1.00} \\ \cline{3-8}
															
						\multicolumn{2}{|c|}{}  &	 \multirow{5}*{\footnotesize FP} & {\footnotesize c=1/16} & {\footnotesize 0.78} & {\footnotesize 0.70} & {\footnotesize 1.00} & {\footnotesize 1.00} \\ \cline{4-8}
						\multicolumn{2}{|c|}{}  &	 {}						 & {\footnotesize c=2/16} & {\footnotesize 0.66} & {\footnotesize 0.53} & {\footnotesize 1.00} & {\footnotesize 1.00} \\ \cline{4-8}
						\multicolumn{2}{|c|}{}  &	 {}						 & {\footnotesize c=3/16} & {\footnotesize 0.59} & {\footnotesize 0.42} & {\footnotesize 1.00} & {\footnotesize 1.00} \\ \cline{4-8}
						\multicolumn{2}{|c|}{}  &	 {}						 & {\footnotesize c=4/16} & {\footnotesize 0.48} & {\footnotesize 0.28} & {\footnotesize 1.00} & {\footnotesize 1.00} \\ \cline{4-8}
						\multicolumn{2}{|c|}{}  &	 {}						 & {\footnotesize c=5/16} & {\footnotesize 0.55} & {\footnotesize 0.20} & {\footnotesize 0.99} & {\footnotesize 0.99} \\ \hline
															
\multirow{6}*{\makecell {\footnotesize Negative \\ \footnotesize suspect \\ \footnotesize classifiers}} 
						& {\makecell{\footnotesize Same \\\footnotesize architecture \\ \footnotesize DNNs}}
							&  \multicolumn{2}{c|}{\footnotesize WRN-22-4} & {\footnotesize [0.02,0.36]} & {\footnotesize [0.00,0.04]} & {\footnotesize [0.00,0.01]} & {\footnotesize [0.00,0.02]}\\\cline{2-8}
					       ~ &  {\makecell{\footnotesize Different \\\footnotesize architecture}} 
							 &  \multicolumn{2}{c|}{\footnotesize LeNet-5} & {\footnotesize [0.00,0.07]} & {\footnotesize [0.00,0.01]} & {\footnotesize [0.00,0.00]} & {\footnotesize [0.00,0.01]}\\ \cline{3-8}
					       ~ &    {\footnotesize DNNs} &  \multicolumn{2}{c|}{\footnotesize VGG16} & {\footnotesize [0.00,0.11]} & {\footnotesize [0.00,0.01]} & {\footnotesize [0.00,0.00]} & {\footnotesize [0.00,0.01]}\\ \cline{2-8}
					       ~ & \makecell{\footnotesize Random\\\footnotesize forests} & \multicolumn{2}{c|}{\footnotesize RF} 
  						      & {\footnotesize [0.00,0.04]} & {\footnotesize [0.00,0.02]} & {\footnotesize [0.00,0.00]} & {\footnotesize [0.00,0.00]}\\ \hline

\end{tabular}
}
\label{matchingrate-CIFAR100}
\end{table}

\begin{table}[!t]\renewcommand{\arraystretch}{1.0}
\centering
\caption{Matching rates of the compared methods for positive and negative suspect classifiers on ImageNet.}
\centering
\addtolength{\tabcolsep}{-2.5pt}
\scalebox{1.0}{
\begin{tabular} {|c|c|c|c|c|c|c|c|}\hline 
\multicolumn{4}{|c|}{\footnotesize Suspect classifier} & {\footnotesize FGSM} & {\footnotesize IGSM} & {\footnotesize CW-$L_2$}  & {\footnotesize IPGuard}\\ \hline

\multicolumn{2}{|c|}{\multirow{10}*{\makecell{\footnotesize Positive \\\\\footnotesize suspect \\\\\footnotesize classifiers}}}
						    	&	\multicolumn{2}{c|}{\footnotesize FTLL} 							& {\footnotesize 1.00} & {\footnotesize 0.95} & {\footnotesize 1.00} & {\footnotesize 1.00} \\ \cline{3-8}
						\multicolumn{2}{|c|}{}  & \multicolumn{2}{c|}{\footnotesize FTAL} 		&{\footnotesize 0.13} & {\footnotesize 0.39} & {\footnotesize 1.00} & {\footnotesize 1.00} \\ \cline{3-8}
						\multicolumn{2}{|c|}{}  &	 \multicolumn{2}{c|}{\footnotesize RTLL} 		& {\footnotesize 0.75} & {\footnotesize 0.32} & {\footnotesize 1.00} & {\footnotesize 1.00} \\ \cline{3-8}
						\multicolumn{2}{|c|}{}  &	 \multicolumn{2}{c|}{\footnotesize RTAL} 		& {\footnotesize 0.43} & {\footnotesize 0.10} & {\footnotesize 0.99} & {\footnotesize 0.99} \\ \cline{3-8}
						\multicolumn{2}{|c|}{}  & \multirow{5}*{\footnotesize WP} & {\footnotesize p=0.1} & {\footnotesize 0.96} & {\footnotesize 0.83} & {\footnotesize 1.00} & {\footnotesize 1.00} \\ \cline{4-8}
						\multicolumn{2}{|c|}{}  & {}						 & {\footnotesize p=0.2} & {\footnotesize 0.65} & {\footnotesize 0.81} & {\footnotesize 1.00} & {\footnotesize 1.00} \\ \cline{4-8}
						\multicolumn{2}{|c|}{}  & {}						 & {\footnotesize p=0.3} & {\footnotesize 0.58} & {\footnotesize 0.77} & {\footnotesize 1.00} & {\footnotesize 1.00} \\ \cline{3-8}
						
						\multicolumn{2}{|c|}{}  &	 \multirow{3}*{\footnotesize FP} & {\footnotesize c=1/16} & {\footnotesize 0.82} & {\footnotesize 0.02} & {\footnotesize 1.00} & {\footnotesize 0.99} \\ \cline{4-8}
						\multicolumn{2}{|c|}{}  &	 {}						 & {\footnotesize c=2/16} & {\footnotesize 0.01} & {\footnotesize 0.02} & {\footnotesize 1.00} & {\footnotesize 0.96} \\ \cline{4-8}
						\multicolumn{2}{|c|}{}  &	 {}						& {\footnotesize c=3/16} & {\footnotesize 0.00} & {\footnotesize 0.01} & {\footnotesize 0.99} & {\footnotesize 0.90} \\ \hline

\multirow{10}*{\makecell {\footnotesize Negative \\\\ \footnotesize suspect \\\\ \footnotesize classifiers}} 
					       ~ &  \multirow{10}*{\makecell{\footnotesize Different\\\\ \footnotesize architecture\\\\\footnotesize DNNs}} 
							 & \multicolumn{2}{c|}{\footnotesize VGG16} 						& {\footnotesize 0.32} & {\footnotesize 0.00} & {\footnotesize 0.00} & {\footnotesize 0.01}\\ \cline{3-8}
					       ~ &    ~ & \multicolumn{2}{c|}{\footnotesize ResNet152} 					& {\footnotesize 0.02} & {\footnotesize 0.00} & {\footnotesize 0.00} & {\footnotesize 0.00}\\ \cline{3-8}
					       ~ &    ~ & \multicolumn{2}{c|}{\footnotesize ResNet152V2}				& {\footnotesize 0.00} & {\footnotesize 0.00} & {\footnotesize 0.00} & {\footnotesize 0.00}\\ \cline{3-8}
					       ~ &    ~ & \multicolumn{2}{c|}{\footnotesize InceptionV3} 				& {\footnotesize 0.00} & {\footnotesize 0.00} & {\footnotesize 0.00} & {\footnotesize 0.00}\\ \cline{3-8}
					       ~ &    ~ & \multicolumn{2}{c|}{\footnotesize InceptionResNetV2} 			& {\footnotesize 0.00} & {\footnotesize 0.00} & {\footnotesize 0.01} & {\footnotesize 0.00}\\ \cline{3-8}
					       ~ &    ~ & \multicolumn{2}{c|}{\footnotesize Xception} 					& {\footnotesize 0.00} & {\footnotesize 0.00} & {\footnotesize 0.00} & {\footnotesize 0.00}\\ \cline{3-8}
					       ~ &    ~ & \multicolumn{2}{c|}{\footnotesize MobileNet($\alpha$=1.0)} 	& {\footnotesize 0.00} & {\footnotesize 0.00} & {\footnotesize 0.01} & {\footnotesize 0.00}\\ \cline{3-8}
					       ~ &    ~ & \multicolumn{2}{c|}{\footnotesize MobileNetV2($\alpha$=1.4)} 	& {\footnotesize 0.00} & {\footnotesize 0.00} & {\footnotesize 0.01} & {\footnotesize 0.00}\\ \cline{3-8}
					       ~ &    ~ & \multicolumn{2}{c|}{\footnotesize DenseNet201} 				& {\footnotesize 0.00} & {\footnotesize 0.00} & {\footnotesize 0.04} & {\footnotesize 0.00}\\ \cline{3-8}
					       ~ &    ~ & \multicolumn{2}{c|}{\footnotesize NASNetLarge} 				& {\footnotesize 0.00} & {\footnotesize 0.00} & {\footnotesize 0.00} & {\footnotesize 0.00}\\  \hline

\end{tabular}
}
\label{matchingrate-ImageNet}
\end{table}

\myparatight{Matching rate} Table~\ref{matchingrate-CIFAR10}, Table~\ref{matchingrate-CIFAR100}, and Table~\ref{matchingrate-ImageNet} respectively show the matching rates of the compared methods for the three datasets, where the parameter of each method is set such that it achieves the best ARUC on a dataset. For each category of negative suspect classifiers, we have multiple instances. Therefore, we show the range of matching rates for them. We have several observations. 

First,  positive suspect classifiers have larger matching rates than negative suspect classifiers in most cases. In particular, for IGSM, CW-$L_2$, and IPGuard, there exists proper matching rate thresholds that make both robustness and uniqueness equal to 1.  These results may indicate that the classification boundary of a target classifier is closer to those of the post-processed versions of the target classifier than those of other classifiers. 
 However, CW-$L_2$ and IPGuard achieve the largest gaps between matching rates of positive suspect classifiers and those of negative suspect classifiers. In other words, for a wider range of matching rate threshold, CW-$L_2$ and IPGuard can achieve both robustness and uniqueness of 1. For instance, on CIFAR10, IGSM, CW-$L_2$, and IPGuard respectively achieve  gaps of 0.06, 0.03, 0.63, and 0.62 between the smallest matching rate of positive suspect classifiers and the largest matching rate of negative suspect classifiers. Our evaluated suspect classifiers are a subset of suspect classifiers in practice. A larger such gap makes it more likely to select a matching rate threshold that achieves both large robustness and uniqueness in practice.   

Second, two neural network classifiers, which use the same architecture but different initializations when learning the model parameters, have substantially different classification boundaries, even if their testing accuracies are close. For instance, on CIFAR-100, the target classifier is a WRN-22-4 model; and the other WRN-22-4 models that are trained with different initializations have matching rates that are close to 0 for the compared methods except FGSM.   Third, among the post-processing techniques, FP tends to have lower matching rates, which indicates that FP changes the target classifier's classification boundary more substantially. 

\myparatight{Efficiency} Table~\ref{tab:efficiency} shows the running time of generating the 100 fingerprinting data points for different methods, where the parameter of each method is set to achieve the best ARUC on a dataset. IPGuard is slower than FGSM and IGSM. However, IPGuard is orders of magnitude faster than CW-$L_2$.   IPGuard and CW-$L_2$ achieve comparable ARUCs, but IPGuard is more efficient than CW-$L_2$. This is because they both can find data points near a target classifier's classification boundary, but IPGuard does not constrain noise added to the data points to be more efficient.    
For instance, on the CIFAR-100 dataset, the average $L_2$-norms of the noise added by CW-$L_2$ and IPGuard to the initial data points are 0.64 and 1.29, respectively. 

\begin{table}[!t]\renewcommand{\arraystretch}{1.00}
\centering
\caption{Running time (in seconds) of generating the 100 fingerprinting data points for the compared methods.}
\centering
\scalebox{1.0}{
\begin{tabular} {|c|c|c|c|}\hline 
{\small } & {\small CIFAR-10} & {\small CIFAR-100} & {\small ImageNet}\\ \hline
{\small Random} & {\small $<$1} & {\small $<$1} & {\small $<$1}\\ \hline
{\small FGSM} & {\small 4.9} & {\small 4.6} & {\small 11.6}\\ \hline
{\small IGSM} & {\small 11.2} & {\small 15.6} & {\small 47.9}\\ \hline
{\small CW-$L_2$} & {\small{20,006.3}} & {\small 30,644.6} & {\small 121,955.7}\\ \hline
{\small IPGuard} & {\small 37.8} & {\small 249.9} & {\small 7,634.3}\\ \hline
\end{tabular}
}
\label{tab:efficiency}
\end{table}


\section{Discussion and Limitations}

We acknowledge that the effectiveness of IPGuard relies on the assumption that we can find some data points such that a target classifier and its post-processed versions are more likely to predict the same label for each of them, but classifiers that are not post-processed versions of the target classifier are more likely to predict different labels for them. In other words, IPGuard relies on the assumption that the classification boundary of a target classifier is closer to those of its post-processed versions than to those of other classifiers. Our empirical results show that this assumption is valid for popular post-processing techniques including fine tuning, retraining, and model compression. We did not evaluate model extraction attacks~\cite{tramer2016stealing,WangHyper18,juuti2018prada,oh2017towards,hua2018reverse, yan2018cache,Hu19} due to limited resources and we consider it as an interesting future work. 

We envision that there is an arms race between a model owner and an attacker. An attacker could strategically  post-process the target classifier to evade the model owner's fingerprinting method, while a model owner aims to  fingerprint a target classifier such that any post-processed version of its target classifier can be identified. Specifically, an attacker can find some way to post-process the model such that the classification boundary changes significantly without sacrificing the classification accuracy too much. One possible adaptive attack may be post-processing the model via knowledge distillation. We note that, once stealing and post-processing a target classifier requires less resource (e.g., computation resource, training data) than training a classifier from scratch, an attacker may be motivated to steal and/or post-process the target classifier instead of training a classifier from scratch. Therefore, the arms race between a model owner and an attacker may last until  stealing the target classifier and evading the model owner's fingerprinting method requires  more resource than training a classifier from scratch. One way to end the arms race is to provide some theoretical guarantees on the fingerprinting method, which is out of the scope of this paper and we leave it as a future work.


\section{Conclusion}

In this work, we propose to leverage data points near a classifier's classification boundary to fingerprint the classifier and track its use in the wild. Moreover, we propose a new method to efficiently find such data points. Our empirical results on three large-scale benchmark datasets show that 1) the fingerprint extracted by our method can be robust against popular post-processing of the classifier and unique to the classifier (and its post-processed versions) simultaneously; and 2) our method is more efficient than state-of-the-art adversarial example methods at finding data points near the classification boundary because our method does not constrain the noise added to such data points. 


\section*{Acknowledgement}
We thank the anonymous reviewers and our shepherd Tianwei Zhang for insightful reviews and comments. This work was supported by NSF grant No.1937786.


\balance{\bibliographystyle{ACM-Reference-Format}
\bibliography{refs}}

\end{document}